\documentclass[final,12pt]{article}

\usepackage[usenames,dvipdfmx]{color}
\usepackage{setspace,amsfonts,comment,amsmath,amsthm,amssymb,amsxtra,graphicx}
\usepackage[pdftex,pagebackref]{hyperref}
\usepackage{mathpazo}

\usepackage{threeparttable}
\usepackage{booktabs}
\usepackage[T1]{fontenc}
\usepackage{inputenc}  
\usepackage{esint}
\usepackage[authoryear]{natbib}
\usepackage{sidecap}
\usepackage[font=footnotesize]{caption}
\usepackage{pgfplots}

\pgfplotsset{width=7cm,compat=1.8}

\usepackage{float}

\hypersetup{pdfpagemode=None, colorlinks=true,
 anchorcolor= webbrown, citecolor= blue,
filecolor= webbrown, linkcolor= blue, menucolor= webbrown,
urlcolor= webbrown, citebordercolor= 1 0 0, menubordercolor=1 0
0, urlbordercolor=1 0 0, runbordercolor=1
0 0 } \hypersetup{pdfauthor=Bruno Levy}
\definecolor{webgreen}{rgb}{.6,.6,.6}
\definecolor{webbrown}{rgb}{.6,0.15,0.15}
\definecolor{webyellow}{rgb}{0.98,0.92,0.73}

\newcommand\ddate{\today}
\newcommand\aauthorA{\footnote{brunopcl@al.insper.edu.br} Bruno P. C. Levy  }
\newcommand\aauthorB{\footnote{hedibertfl@insper.edu.br} Hedibert F. Lopes}
\newcommand\aaffiliationA{Insper}
\newcommand\aaffiliationB{Insper}
\newcommand\eemailA{\href{mailto:brunopcl@al.insper.edu.br}{{\small\texttt{{}}}}}
\newcommand\eemailB{\href{mailto:dguillen@gmail.com}{{\small\texttt{{}}}}}
\newcommand\ttitle{Dynamic Ordering Learning in Multivariate Forecasting}
\newcommand\tthanks{We thank to all participants at Insper Seminars for useful comments. All remaining errors are of our responsibility.
}

\newcommand{\propref}[1]{\hyperref[#1]{Proposition~\ref*{#1}}}
\newcommand{\appref}[1]{\hyperref[#1]{Appendix~\ref*{#1}}}
\newcommand{\secref}[1]{\hyperref[#1]{Section~\ref*{#1}}}

\newcommand{\commentout}[1]{}

\newtheorem*{lemma*}{Lemma}

\hypersetup{pdfstartview=FitH}
\title{\vspace{-1.5\baselineskip}  
{\ttitle\thanks{~\tthanks}}\bigskip}

\author{{\aauthorA} \\   
{\small \aaffiliationA}\\
[-.1in] \eemailA \and
{\aauthorB} \\       
{\small \aaffiliationB}\\
[-.1in] \eemailB \medskip }
\date{\normalsize \textcolor{webgreen}{\ddate}}
\date{\normalsize \textcolor{webgreen}{Working Paper - November, 2021}}

\begin{document}

\maketitle

\vspace{-1.5\baselineskip}

\begin{abstract} 

In many fields where the main goal is to produce sequential forecasts for decision-making problems, the good understanding of the contemporaneous relations among different series is crucial for the estimation of the covariance matrix. In recent years, the modified Cholesky decomposition appeared as a popular approach to covariance matrix estimation. However, its main drawback relies on the imposition of the series ordering structure. In this work, we propose a highly flexible and fast method to deal with the problem of ordering uncertainty in a dynamic fashion with the use of Dynamic ordering Probabilities. We apply the proposed method in two different forecasting contexts. The first is a dynamic portfolio allocation problem, where the investor is able to learn the contemporaneous relationships among different currencies improving final decisions and economic performance. The second is a macroeconomic application, where the econometrician can adapt sequentially to new economic environments, switching the contemporaneous relations among macroeconomic variables over time.


\vspace{1cm}


\vspace{1cm}

\noindent\noindent \emph{J.E.L. codes: C11, C52, G11, G17, F31 } 

\end{abstract}

\onehalfspace

\newpage

\tableofcontents

\newpage

\section{Introduction} \label{sec1: Introduction}

Model uncertainty is a well-known challenge among applied researchers and industry practitioners that are interested in producing forecasts for decision-making problems. Such applications range from forecasting macroeconomic series (GDP, inflation, unemployment, interest rates, etc.) to  commercial sales, financial series for portfolio decisions and many others. The fact is that regardless of the field of interest, the decision maker is often uncertain about which model specifications will produce higher forecasting performance, refining final decisions. Common uncertainties are not just about the best predictors to choose, they are also related to the dynamic of the coefficients. Since the environment of the economy and financial markets are always changing over time, an additional important uncertainty for prediction modeling is the dynamic of coefficients. The relationships among economic variables can move in different manners over time, in a higher or lower intensity degree. Some variables respond to others more intensely during some periods and others evolve almost constantly.

Recently, within the Bayesian literature on multivariate forecasting,  the Cholesky-style  began to appear as a natural approach for covariance matrix estimation (\citealp{primiceri2005time}, \citealp*{zhao2016dynamic}, \citealp*{lopes2016parsimony}, \citealp{west2020bayesian}, \citealp*{kastner2020sparse}, \citealp*{fisher2020optimal} and \citealp*{lavine2020adaptive}). The Cholesky-decomposition for covariance matrices guarantees the positive definiteness and allows for higher model flexibility compared to the traditional Inverse-Wishart distribution, since the estimation of different variable volatilities and co-volatilities can now be disentangled and dealt in parallel. However, despite the Cholesky-style's popularity gain, its structure requires the econometrician/researcher to impose a specific - and, in many cases, arbitrary - series ordering. Hence, we argue here that the series ordering can be viewed now as a new possible uncertainty in the modeling process. Therefore, the main goal of our paper is to provide a dynamic approach to sequentially deal with the problem of series \textit{ordering uncertainty} in multivariate forecasting problems.

In the last decades, the Bayesian literature has addressed the question of general model uncertainty with great success. Bayesian Model Averaging (BMA) and Bayesian Model Selection (BMS) are well-known methodologies for static models when there is uncertainty about the predictors to include (\citealp{madigan1994model} and \citealp{hoeting1999bayesian}). More recently, \cite*{raftery2010online} proposed the dynamic version of BMA, called Dynamic Model Averaging (DMA) and Dynamic Model Selection (DMS). In this framework, it is considered the idea that the relevant model can change over time. For instance, a subset of the relevant predictors during the 1980s might become irrelavant during the 1990s, or during the Great Recession or even during the current Covid-19 Crisis. This is a challenging task and even a simple forecasting problem can become computationally infeasible; for instance a researcher choosing among $p$ predictors would have to consider $2^p$ different models at each time period. \cite{raftery2010online} suggest the use of reasonable approximations, borrowing ideas from discount (forgetting) methods, avoiding simulations of transition probabilities matrices and maintaining the conjugate form of posterior distributions, allowing analytical solutions for forward filtering and forecasting which significantly reduces the computation burden of the process.

After \cite{raftery2010online}, several papers appeared in many areas applying DMA and DMS methods. Just to name a few, \cite{koop2012forecasting} were one of the first to use this ideas in a macroeconomic context, forecasting US inflation. \cite{dangl2012predictive} forecast the $S\&P$ 500 index using traditional financial predictors, finding that
time-varying parameter models are preferable to regressions with constant coefficients.  More recently, \cite*{catania2019forecasting}  applied those methods to forecast different cryptocurrencies and, within a dynamic binary classification context, \cite{levy2021time} show significant portfolio performance improvements for investors who sequentially learn different time-series momentum speed combinations for trend-following strategies in future markets.

The literature has also produced examples in the multivariate context, using dynamic model probabilities. Such examples appeared in \cite{koop2013large}, where the authors propose a method to  sequentially learn the best dimension of time-varying parameters VARs. \cite{koop2014new} use in the construction of a financial condition index, with time-varying weights for each financial variable in the index. Recently, \cite{beckmann2020exchange} show statistical and economic improvements for a portfolio allocation problem using TVP-VARs. What is common for the papers on the multivariate case is that they rely on the use of inverse Wishart distribution for the volatility matrix. This Wishart Dynamic Linear Models (W-DLM) is well documented in the Bayesian literature (see \citealp{west2006bayesian}). However, the W-DLM comes with some constraints. It not only imposes all equations in the system to share the same predictors, but also ties the behavior of volatilities and co-volatilities, since they are jointly modeled, avoiding specific customizations.       

In the last couple of years, it began to flourish in the statistic and econometric literature what started to be known as a "Decouple/Recouple" concept, which is built on recursive systems. The basic idea is to decouple the multivariate dynamic model into several univariate customized DLMs and then recouple for forecasting and decisions. It is strictly related to the popular Cholesky-style Multivariate Stochastic Volatility of \cite{lopes2016parsimony},   \cite*{shirota2017cholesky} and \cite{primiceri2005time}. Using similar ideas and extending  multi-regression dynamic models applied in \cite*{queen2008forecast} and \cite{costa2015searching},  
the work of \cite*{zhao2016dynamic} introduced the concept of Dynamic Dependency Network models (DDNM), allowing the use of customized additional predictors and enabling sequential analysis of decoupled univariate DLMs and then recoupled for forecasting. This type of approach has been discussed in \cite{west2020bayesian} and applied in a similar fashion in \cite{lavine2020adaptive} and  \cite{fisher2020optimal} .The huge advantage of DDNM compared to the W-DLM approach is its flexibility to feature its own set of predictors for each equation and enabling volatilities and co-volatilities to vary with different degrees over time. Also, as in the W-DLM, the DDNM yields closed-form solutions for posterior distributions and predictive densities. Hence, there is no need for expensive simulations methods which makes the process much faster.

As we have argued before, even though DDNM introduces flexibility in the process by its closed-format nature and the use of the Cholesky-style framework, it still imposes a specific series ordering in the model (more details will be discussed in Section \ref{sec: Dynamic Dependency Network Models}). The fact is that all models within the Cholesky-style framework will also depend on the ordering structure selected by the researcher, leading to different contemporaneous relations among series. In our work, we argue that this imposition can change the covariance matrix estimation and substantially modify final decisions. Therefore, the researcher in general will face the problem of ordering Uncertainty, where the "correct" ordering structure and contemporaneous relations among variables are unknown.

The problem of ordering uncertainty was recognized by the literature (\citealp{primiceri2005time}, \citealp{zhao2016dynamic} and \citealp*{lopes2016parsimony}), but was not fully solved. Possible paths have been applied, such as proposing some few different orderings and comparing the results among them or even do not impose any Cholesky-style structure and use more approximations and expensive simulations to obtain predictive densities (\citealp{gruber2016gpu}). Recent studies also try to consider the issue in the static formulation (\citealp{zheng2017cholesky} or  \citealp*{kang2020cholesky}), however it does not take into account the fact that the contemporaneous relations among time series may evolve over time, i.e., there is a single ordering structure for all periods of time and parameters are considered constant throughout.

We agree that, for some specific cases, the researcher may know in advance the contemporaneous relations among time series. From macroeconomics, for example, the researcher is able to build the ordering structure borrowing ideas from some theory well developed in the literature. However, for many different cases, the econometrician is faced with a new problem in which it is not totally clear how the data is organized. Or even more, we argue here that this structure is unstable over time. Hence, a common structure in the 1980s can differ from the structure during the Great recession. Also, that latter ordering can be different from nowadays. Therefore, the contemporaneous relations among different variables may change over time, being stronger, weaker or nonexistent depending on the economical environment. Considering the absence of studies dealing with the problem of ordering uncertainty in a dynamic fashion, we propose the {\em Dynamic Ordering Learning} (DOL) approach. Our DOL scheme is a faster and a more flexible method do deal with the uncertainty about the contemporaneous relations among dependent variables and predictors in a online learning environment.  By using the same structure as DDNM and following similar path as \cite{raftery2010online} and \cite{koop2013large} to deal with model uncertainty,  predictive densities have closed-form solutions, therefore avoiding the use of MCMC schemes and substantially reducing the computational burden of the estimation and forecasting processes.

We propose a dynamic method to deal with the uncertainty around series ordering and different contemporaneous dependencies across series. We show in a dynamic portfolio study with exchange rates that our DOL approach generates superior statistical and economic performances for a mean-variance investor that uses the predictive information to dynamically rebalance her portfolio. That is, with DOL the investor is able to sequentially learn the dynamic contemporaneous relations among different currencies and improve the predicted means and covariances of returns over time. We show that the investor will be willing to pay a considerable fee to switch from the traditional Wishart-Random Walk method and fixed orderings over time to our DOL approach. 

Finally, we highlight that the DOL can be applied in any field where the the main goal is to produce sequential forecasts in decision-making problems, such as in Central Banks, financial institutions, commercial industries and others. It motivates our study to verify a second application. We consider the problem of forecasting a set of important macroeconomic series commonly used in the literature: inflation, unemployment and interest rates. We show that the econometrician who learns the changes in the contemporaneous relations among different economic variables is able to improve point and density forecasting compared to the econometrician that considers a single ordering structure for all periods of time. Therefore, it gives evidences that the environment of the economy is continuously changing, raising the importance for forecasters and decision makers to incorporate this dynamic behavior in their econometric models.

The remainder of the paper is organized as follows. In Section \ref{sec: Dynamic Dependency Network Models} we start by introducing the general econometric framework behind the Dynamic Dependency Network model, with a brief discussion about the Cholesky-style approach. In Section \ref{ref: Dynamic Ordering Learning} we introduce and detail the ordering uncertainty problem and our Dynamic Ordering Learning mechanism within the Cholesky-style and how it can be applied to decision-making problems. Section \ref{simulation} provides a simulation study and, in Sections \ref{ref: Portfolio_section} and \ref{ref: Macro_section}, we perform the empirical econometric applications, where the first is related to a portfolio allocation problem and the second to a macroeconomic forecasting context. Finally, Section \ref{sec: Conclusion} concludes.

\section{The Dynamic Dependency Network Model}	
\label{sec: Dynamic Dependency Network Models}

As previously mentioned, the DDNM framework of \cite{zhao2016dynamic} is able to model cross-section contemporaneous relationships and customize univariate DLMs. To make it clear, we begin by revisiting its modeling framework for a single and arbitrary ordering structure and, in Section (\ref{ref: Dynamic Ordering Learning}), we extend the approach to incorporate the whole ordering uncertainty mechanism. 

First, consider $\boldsymbol{y_{t}}$ as a $\textit{m}$-dimensional vector with (financial/economic) time series $y_{j,t}$ and consider the following dynamic system: 

\vspace{4mm}

\begin{equation}
\left(\boldsymbol{I}_m-\boldsymbol{\Gamma}_{t}\right)
\boldsymbol{y}_{t}=\left(\begin{array}{c}
\boldsymbol{x}_{1, t-1}^{\prime} \boldsymbol{\beta}_{1 t} \\
\vdots \\
\boldsymbol{x}_{m, t-1}^{\prime} \boldsymbol{\beta}_{m t}
\end{array}\right)
+ \boldsymbol{\nu_{t}},
\quad \quad \boldsymbol{\nu_{t}} \mid \boldsymbol{\Omega}_{t} \sim \mathcal{N}\left(\boldsymbol{0}, \boldsymbol{\Omega}_{t}\right),
\label{eq:ddnm}
\end{equation}

\vspace{4mm}

\noindent where $\boldsymbol{x}_{j, t-1}$ is a $\textit{p}$-dimensional vector of time series j's lag predictors, $\boldsymbol{\beta}_{j t}$ are time-varying coefficients and  $\boldsymbol{\Omega}_{t} = \operatorname{diag}
\left(\sigma_{1 t}^{2}, \ldots, \sigma_{m t}^{2}\right)$.
Therefore, all the contemporaneous relations among time series are coming from the $\textit{m}$ $\times$ $\textit{m}$ matrix $\boldsymbol{\Gamma}_{t}$, whose off-diagonal elements $\gamma_{ji t}$s (for $j \neq i $) capture the dynamic contemporaneous relationships among series $j$ and $i$ at time $t$.
$\boldsymbol{\Gamma}_{t}$ has zeroes on the main diagonal.


Throughout our work, and following 
\cite{zhao2016dynamic}, we will focus on the particular but important case where $\boldsymbol{\Gamma_{t}}$ is lower triangular with zeroes in and above the main diagonal:
\vspace{4mm}

\begin{equation}
\boldsymbol{\Gamma}_t=\left[\begin{array}{ccccc}
0 & 0 & \ldots & 0 & 0 \\
\gamma_{21, t} & 0 & \ldots & 0 & 0 \\
\vdots & \vdots & \ddots & \vdots & \vdots \\
\gamma_{m 1, t} & \gamma_{m 2, t} & \ldots & \gamma_{m, m-1, t} & 0
\end{array}\right]
\label{eq:triangular}
\end{equation}

\vspace{4mm}

This particular case has already appeared in the econometric literature.  \cite{lopes2016parsimony}, for example, deal with time-varying learning of covariance matrices with no predictors and handle hundreds of time series simultaneously via parsimonious priors (see also \cite*{shirota2017cholesky}).  Also, \cite{primiceri2005time} uses lagged values of $\boldsymbol{y}_{t}$ in a VAR with stochastic volatility context with random walk dynamics for $\boldsymbol{\beta}_{it}$s, $\gamma_{ijt}$s and $\sigma_{it}^2$s.


Since the error terms in  $\boldsymbol{\nu_{t}}$ are contemporaneouly uncorrelated, the triangular contemporaneous dependencies among time series in Equation (\ref{eq:triangular}) generate a fully recursive system, known as a Cholesky-style framework (\citealp{west2020bayesian}). Hence, each equation $j$ of the system will have its own set of $\textit{parents}$ ($\boldsymbol{y_{<j, t}}$), that is, will depend contemporaneously on all other time series above equation $j$, following the triangular format in Equation (\ref{eq:triangular}). In words, the top time series in the system will not have parents, the second from the top time series will have the first time series as a parent, the third time series will have the first two time series as parents all the way to the last time series, which will depend on all other $m-1$ time series above it. This triangular form makes clear the understanding of how the imposition of the ordering structure implies a specific contemporaneous dependency among time series.

Equation (\ref{eq:ddnm}) can be rewritten in the reduced form as
\begin{equation}
\boldsymbol{y}_{t}=\boldsymbol{A}_t\left(\begin{array}{c}
\boldsymbol{x}_{1, t-1}^{\prime} \boldsymbol{\beta}_{1 t} \\
\vdots \\
\boldsymbol{x}_{m, t-1}^{\prime} \boldsymbol{\beta}_{m t}
\end{array}\right)+\boldsymbol{u}_{t} \quad \boldsymbol{u}_{t} \mid \boldsymbol{\Sigma}_{t} \sim \mathcal{N}\left(0, \boldsymbol{\Sigma}_{t}\right)
\label{eq:reducedform}
\end{equation}
where $\boldsymbol{A}_t=\left(\boldsymbol{I}_m-\boldsymbol{\Gamma}_{t}\right)^{-1}$ and $\boldsymbol{u}_t = \boldsymbol{A}_t \boldsymbol{\nu_{t}}$. The modified Cholesky decomposition clearly appears in $\boldsymbol{\Sigma}_{t}=\boldsymbol{A}_t \boldsymbol{\Omega}_{t}\boldsymbol{A}_t^{\prime}$ which is now a full variance-covariance matrix capturing the contemporaneous relations among the $\textit{m}$ time series. Given the parental triangular structure of $\boldsymbol{\Gamma_{t} }$ in (\ref{eq:triangular}), the equations will be conditionally independent, bringing the ``Decoupled'' aspect of the multivariate model. In other words, the DDNM can be viewed as a set of $\textit{m}$ conditionally independent univariate DLMs that can be dealt with in a parallelizable fashion. The outputs of each equation are then used to compute $\boldsymbol{\Gamma_{t}}$ and $\boldsymbol{\Omega_{t}}$, hence recovering the full time-varying covariance matrix $\boldsymbol{\Sigma_{t}}$. 

\subsection{$\textit{m}$ univariate dynamic linear models}
The set of $m$ univariate models can be represented as $m$ univariate recursive dynamic regressions, for $j = 1, \dots, m$:

\begin{equation}
y_{j t}=\boldsymbol{x}_{j, t-1}^{\prime} \boldsymbol{\beta}_{j t}+\boldsymbol{y}_{<j, t}^{\prime} \boldsymbol{\gamma}_{<j, t}+\nu_{j t},  \quad \quad \nu_{j t} \sim \mathcal{N}\left(0, \sigma_{j t}^{2}\right),
\label{eq:dlms}
\end{equation}
and dynamic coefficients evolving according to random walks:
\begin{equation}
\left(\begin{array}{c}
\boldsymbol{\beta}_{j t} \\
\boldsymbol{\gamma}_{<j, t}
\end{array}\right)=\left(\begin{array}{c}
\boldsymbol{\beta}_{j, t-1} \\
\boldsymbol{\gamma}_{< j, t-1}
\end{array}\right)+\boldsymbol{\omega}_{j t} \quad \boldsymbol{\omega}_{j t} \sim \mathcal{N}\left(\boldsymbol{0}, \boldsymbol{W}_{j t}\right).
\end{equation}
By defining the full dynamic state and regression vectors as
$$
\boldsymbol{\theta}_{j t}=\left(\begin{array}{c}\boldsymbol{\beta}_{j t} \\ \boldsymbol{\gamma}_{< j, t}\end{array}\right) \quad \text{and} \quad \mathbf{F}_{j t}=\left(\begin{array}{c}\boldsymbol{x}_{j, t-1} \\ \boldsymbol{y}_{<j,t}\end{array}\right),
$$
we recover the traditional conditionally linear and gaussian univariate DLM formulation as in \cite{west2006bayesian}, namely
\begin{eqnarray*}
y_{j t} &=& \mathbf{F}_{j t}^{\prime} \boldsymbol{\theta}_{j t}+\nu_{j t},  \quad \quad \nu_{j t} \sim N\left(0, \sigma_{j t}^{2}\right),\\
\boldsymbol{\theta}_{j t} &=& \boldsymbol{\theta}_{j,t-1}+\boldsymbol{\omega}_{j t},  \quad \quad \boldsymbol{\omega}_{j t} \sim N\left(\boldsymbol{0}, \boldsymbol{W}_{j t}\right),
\end{eqnarray*}
for $j=1,\ldots,m$, where again the evolution of $\beta_{jt}$ and $\gamma_{jt}$ evolve over time as a simple random-walk.

\paragraph{Posterior at $t-1$.} Let $\mathcal{D}_{t-1}$ represents the whole information set available until time $t-1$. Then, following the algorithmic structure of sequential learning in DLMs \cite{west2006bayesian}, Chapter 4, at time $t-1$ and for each time series $j$, the joint posterior distribution of $\boldsymbol{\theta}_{j t-1} $ and $\sigma_{j t-1}$ at time $t-1$ can be represented as a multivariate Normal-Gamma\footnote{In both the portfolio allocation and the macroeconomic forecasting studies we center the initial states on the specific models' OLS estimates over the training dataset. Hence, $s_{j, 0}$ is set to be the sample variance estimate of residuals and $m_{j, 0}$ the vector of estimated coefficients. We also set an uninformative state covariance matrix, $C_{j,0} = 100I$, and $n_{j,0} = 10$.  }:

\begin{equation}
\boldsymbol{\theta}_{j, t-1}, \sigma_{j t-1}^{-1} \mid \mathcal{D}_{t-1} \sim \mathcal{NG}\left(\boldsymbol{m}_{j, t-1}, \boldsymbol{C}_{j, t-1}, n_{j, t-1}, n_{j, t-1} s_{j, t-1}\right).
\label{eq:postt1}
\end{equation}

\noindent representing the conditional Normal and marginal Gamma distributions

$$
\theta_{j t-1} \mid \sigma_{j t-1}^{2}, \mathcal{D}_{t-1} \sim \mathcal{N}\left(m_{j t-1}, \frac{\sigma_{j t-1}^{2}}{s_{j t-1}} C_{j t-1}\right)
$$

$$
\sigma_{j t-1}^{-2} \mid \mathcal{D}_{t-1} \sim \mathcal{G}\left(\frac{n_{j t-1}}{2}, \frac{n_{j t-1} s_{j t-1}}{2}\right)
$$

\noindent 
Through the random walk evolution and conjugacy, we can derive the joint prior distribution of $\boldsymbol{\theta}_{jt} $ and $\sigma_{jt}$ for time $t$ as: 
\begin{equation}
\boldsymbol{\theta}_{jt}, \sigma_{jt}^{-1} \mid \mathcal{D}_{t-1} \sim\mathcal{NG}\left(\boldsymbol{a}_{j t}, \boldsymbol{R}_{j t}, r_{j t}, r_{j t} s_{j, t-1}\right)
\end{equation}
where $r_{j t}=\kappa_{j} n_{j, t-1}$, $\boldsymbol{a}_{j t} = \boldsymbol{m}_{j, t-1}$ and  $\boldsymbol{R}_{j t} = \boldsymbol{C}_{j, t-1} / \delta_{j}$.  The quantities $0 < \delta_{j} \leq 1$ and $0  < \kappa_{j} \leq 1$ represent specific discount (aka forgetting) factors for $\theta_{jt}$ and $\sigma_{jt}$, respectively. Discount methods are used to induce time-variations in the evolution of parameters and have been extensively used in many applications (\citealp{raftery2010online}, \citealp{dangl2012predictive}, \citealp{koop2013large}, \citealp*{mcalinn2020multivariate}, amongst others) and well documented in 
\cite{west2006bayesian}, \cite{lopes2006mcmc} and \cite{prado2010time}.

\paragraph{1-step ahead forecast at $t-1$.} The (prior) predictive distribution of $y_{j t}$ is a Student's $t$ distribution with $r_{j t}$ degrees of freedom:
$$
y_{j t} \mid \boldsymbol{y}_{<j,t}, \mathcal{D}_{t-1} \sim \mathcal{T}_{r_{j t}}\left(f_{j t} ,   q_{j t}\right),
$$
with $f_{j t} = \mathbf{F}_{j t}^{\prime}\boldsymbol{a_{j t}}$ and $q_{j t} = s_{j, t-1}+\mathbf{F}_{j t}^{\prime} \mathbf{R}_{j t} \mathbf{F}_{j t} $. It is important to notice that in this framework we have a conjugate analysis for forward filters and one-step ahead forecasting. Therefore, we are able to compute closed-form solution for predictive densities for each equation $j$. Hence, conditional on $\textit{parents}$, it is easy to compute the joint predictive density for $\boldsymbol{y_{t}} $:
\begin{equation}
p\left(\boldsymbol{y}_{t} \mid \mathcal{D}_{t-1}\right)=\prod_{j=1}^m p\left(y_{j t} \mid \boldsymbol{y}_{< j t}, \mathcal{D}_{t-1}\right),
\label{eq:predictive}.
\end{equation}
which simply is the product of the already computed $m$ different univariate Student's $t$ distributions. After the time series are decoupled for sequential analysis, they are then recoupled for multivariate forecasting. In our decision analysis at the empirical Section \ref{ref: Portfolio_section}, we will be concerned with the mean and variance of this distribution for the portfolio allocation study:
\begin{equation}
\boldsymbol{f}_{t}=E\left(\mathbf{y}_{t} \mid \mathcal{D}_{t-1}\right) \ \ \ \mbox{and} \ \ \  \boldsymbol{Q}_{t}=V\left(\mathbf{y}_{t} \mid \mathcal{D}_{t-1}\right).
\end{equation}
Further details about the derivations of the evolution, forecasting, updating distributions can be found in Appendices A and B.

\vspace{4mm}

\section{Dynamic Ordering Learning}
\label{ref: Dynamic Ordering Learning}

In the previous Section, we have discussed the general format of DDNMs using a specific given ordering structure. As we have argued, time series ordering has the potential of being more flexible and more dynamic than simply considering a single structure for all periods of time. The fact is that for a multivariate model with $m$ dependent series, it is possible to have $m!$ different ordering permutations. \cite{raftery2010online} and \cite{koop2013large} have proposed the use of dynamic model probabilities, where the model space was defined by different models with specific predictors and discount factors. Following a similar idea, we propose what we call as {\em Dynamic ordering Probabilities} (DOP), where the model space is set to have models with different ordering structures. Therefore, our model space will contain $m!$ possible orderings. For any order, we dynamically compute probabilities and, conditionally on those and for each period of time, we can select the ordering that received the highest predicted probability or just average the predicted outputs of all orderings weighing by each ordering probability. We name those two approaches {\em Dynamic ordering Selection} (DOS) and {\em Dynamic ordering Averaging} (DOA), respectively. In this way, the econometrician is able to sequentially learn the orderings that have performed well in the recent past and, consequently, learn from past mistakes. We highlight here that the notion of model probabilities has already been applied in the DDNM literature before. For example, \cite{zhao2016dynamic}, \cite{fisher2020optimal} and \cite{lavine2020adaptive} have considered the uncertainty around predictors and discount factors for specific equations in the multivariate system conditioning on a predetermined order. 

Our {\em Dynamic Ordering Learning} (DOL) approach works as follows. Suppose the researcher is faced with the problem of forecasting a multivariate model in which she is uncertain about the ordering structure and does not know exactly the contemporaneous relation between the variables of interest. Additionally, as a refinement, she can also incorporate the case where there are uncertainties around the specification choices around predictors and degrees of variations in parameters for each equation of the system.  We will explain how to incorporate uncertainty around predictors and discount factors in Section \ref{sec: predictors learning}. 

For each ordering $i$, $i = 1, 2, \dots, m! $, we are going to have a set of conditionally independent univariate DLMs with a given parental set structure.  Similar to equation (\ref{eq:predictive}), we can compute the predictive density for each equation $j$ at ordering $i$ and then simply generate the joint  predictive density for ordering $i$ as: 
\begin{equation}
 p\left(\mathbf{y}_{t} \mid \mathcal{D}_{t-1}, \mathcal{O}_{i} \right)=\prod_{j=1}^m p\left(y_{j t} \mid \mathbf{y}_{<j t},
 \mathcal{D}_{t-1}, \mathcal{O}_{i}\right).
 \label{eq:predord}
\end{equation}

Let denote $K = m!$ as the total number of possible orderings. Then, after computing the joint predictive density for all $K$ orderings, we just follow the laws of probability and compute the DOP. First, denoting 

$$
\pi_{t-1 | t-1, i} = Pr(\mathcal{O}_{i} \mid  \mathcal{D}_{t-1}),
$$ 

\noindent as the posterior probability of ordering $i$ at time $t-1$ and following \cite{raftery2010online}, the predicted probability of ordering $i$, given data until time $t-1$ can be expressed as:

\begin{equation}
   \pi_{t | t-1, i}=\frac{\pi_{t-1 | t-1, i}^{\alpha}}{\sum_{l=1}^{K} \pi_{t-1 | t-1, l}^{\alpha}} ,
   \label{eq:pis0}
 \end{equation}

\noindent where $0 < \alpha \leq 1$ is a forgetting factor. The main advantage of using $\alpha$ is avoiding the computational burden associated with expensive MCMC schemes to simulate transitions between orderings over time. This approach has also been extensively used in the Bayesian econometric literaure with great sucess (\citealp{koop2013large}, \citealp{zhao2016dynamic}, \citealp{lavine2020adaptive} and \citealp{beckmann2020exchange}). After observing new data at time $t$, we can compute and use the joint predictive density at Equation (\ref{eq:predord}) to update our ordering probabilities following a simple Bayes' update:

\begin{equation}
    \pi_{t | t, i}=\frac{\pi_{t | t-1, i}p\left(\mathbf{y}_{t} \mid \mathcal{D}_{t-1}, \mathcal{O}_{i} \right)}{\sum_{l=1}^{K} \pi_{t | t-1, l}p\left(\mathbf{y}_{t} \mid \mathcal{D}_{t-1}, \mathcal{O}_{l} \right)},
\label{eq:pis1}
\end{equation}

the posterior probability of ordering $i$ at time $t$. Hence, upon the arrival of a new data point, the researcher is able to measure the performance for each ordering $i$ and to assign higher probability for those orderings that generate better performance. One possible interpretation for the forgetting factor $\alpha$ is through its role to discount past performance. Combining the predicted and posterior probabilities, we can show that 
\begin{equation}
\pi_{t \mid t-1, i} \propto \prod_{l=1}^{t-1}\left[p\left(\mathbf{y}_{t-l} \mid \mathcal{D}_{t-l-1}, \mathcal{O}_{i} \right)\right]^{\alpha^{l}}.
\label{eq:pis}
\end{equation}
Since $0 < \alpha \leq 1$, Equation (\ref{eq:pis}) can be viewed as a discounted predictive likelihood, where past performances are discounted more than recent ones. It implies that orderings that received higher performance in the recent past will produce higher predictive ordering probabilities. The recent past is controlled by $\alpha$, since a lower $\alpha$ discounts more heavily past data and generates a faster switching behavior between orderings over time. Following \cite{beckmann2020exchange}, we induce time-variation in $\alpha$ by considering a grid of values, selecting the best
value for each period of time. In this way, we can allow for periods of faster or slower
ordering switching \footnote{At each time period $t-1$, we select those orderings with the highest predicted probabilities ($\pi_{t \mid t-1, i}$) for each $\alpha$ in the grid. Given those
orderings, we select the alpha that generated the ordering with the highest sum of log predictive likelihood in
the past until time $t-1$. After that, we compute model probabilities based on this best $\alpha$.}.  

After computing ordering probabilities the research is able to deal with the problem of ordering uncertainty by sequentially learning about the importance of each ordering over time. As mentioned before, with predicted ordering probabilities at hand, the researcher can select the ordering that received the highest probability (DOS) or average the predicted outputs of all orderings weighting by each ordering probability (DOA).

One possible limiting aspect of our approach can be seen by its dimensionality restriction. Once there are $m!$ possible orders to be computed, a  large set of time series in $\boldsymbol{y_t}$ would lead to a vast model space, which would dramatically increase the computational burden. For instance, a vector $\boldsymbol{y_t}$ containing 13 variables would generate more than 6 billion possible ordering permutations. Even though our approach allows us to explore the overall set of ordering possibilities only in a restricted dimensionality, we do not see it as a major problem. First, in many cases the econometrician is interested to explore the contemporaneous relations among different subgroups of variables and not necessarily the dependencies among each individual variable. For example, the dynamic dependencies among a large set of asset returns can be much more related to different asset classes/groups (for instance, equities, bonds, commodities, currencies, etc) than in terms of a particular asset movement. Hence, instead of learning the overall set of ordering permutations, the econometrician can explore different asset class orderings, considering a fixed ordering within each asset class and expanding the total set of variables in $\boldsymbol{y_t}$. Second, in many economic problems, the researcher intuition from experience or academic literature can also serve as an important tool to impose contemporaneous dependencies among a subset of variables, leading the ordering uncertainty only to a small/medium scale subgroup of variables. Therefore, we argue here that grouping different variables or incorporating the researcher prior knowledge about a subset of data dependencies can be viewed as a powerful way to circumvent the dimensionality problem. 

\subsection{Predictors and discount factors learning}
\label{sec: predictors learning}

Since the environment of the economy is continuously changing, we apply our DOL approach to sequentially learn the contemporaneous relations among different economic variables over time, improving the covariance matrix estimation. In our application, the decisions about specification choices are quite flexible, allowing not just the econometrician to learn about contemporaneous dependencies, but also about the best predictors and degrees of variations in coefficients and volatilities over time.    

In order to sequentially select the best specification choices for predictors and degree of variation in coefficients, we apply for each equation $j$ (for a given ordering $i$) the Dynamic Model Selection (DMS) approach, similar to what had been done in \cite{raftery2010online}, \cite{koop2012forecasting} and \cite{levy2021time}. The procedure simply selects at each period $t$ the model specification that received the highest predictive model probability. Therefore, an ordering structure will be defined as the joint model with the best univariate model in each equation, respecting the triangular and parental format of that specific order. In our econometric applications, we decided to maintain all specific parents for equation $j$ for a given ordering structure $i$, following the whole triangular format.  


Given the selection of the best univariate model for all equations $j$ at ordering $i$, we can recover the best multivariate model for that ordering structure. To make it clear, consider $\mathcal{M}_{j}^{*}$ the univariate model for equation $j$ at ordering $i$ selected by the DMS procedure at period $t$ with the highest model probability, and let $P(\mathcal{M}_{j}^{*}|\mathcal{D}_{t-1}, \mathcal{O}_{i})$ be its model probability. Since each equation is conditionally independent of the others in the same ordering structure, it imples that   $P(\mathcal{M}_{1:m}^{*}|\mathcal{D}_{t-1}, \mathcal{O}_{i})  = \prod_{j=1}^m P(\mathcal{M}_{j}^{*}|\mathcal{D}_{t-1}, \mathcal{O}_{i})  $, where $\mathcal{M}_{1:m}^{*}$ represents the multivariate model with the highest probability. Hence, as soon as we are able to find the best $m$ univariate models within an ordering structure, we can easily recover the best joint model and compute its joint predictive density and Dynamic Ordering Probabilities, following Equations (\ref{eq:pis0}) and (\ref{eq:pis1}) in Section \ref{ref: Dynamic Ordering Learning}. Therefore, our DOL method proposes considerable flexibility in the specification choices, allowing the econometrician to adapt to new forecasting environments and learning from past mistakes, switching to new predictors, variation in coefficients and volatilities and different contemporaneous dependencies over time.

\section{Simulation Study}
\label{simulation}

The aim of this section is to investigate the performance of our $\textit{Dynamic Ordering Learning}$ approach when applied to a simulated data. Our goal is to give evidences that as soon as a new ordering structure is introduced in the data, our model is able to track the new information in a matter of  few periods and assign higher probability to this new true order. 

We set an artificial data set where ten different series are simulated for 600 data points in time. Suppose that for each hundred time periods, the system changes its dependency structure and a new ordering equivalent to the opposite of the previous one takes place. Therefore, after building a dynamic system that changes its data generated process (DGP) over time we analyze how our model is able to adapt to new environments by looking to the behavior of \textit{Dynamic Ordering Probabilities} (DOP). 

In our simulations we consider the following DGP for a given subperiod of time:

\begin{equation}
y_{jt} =\beta_{j t} y_{j, t-1}+\boldsymbol{y}_{<j, t}^{\prime} \boldsymbol{\gamma}_{<j, t} +\sigma_{t} \varepsilon_{t}, \quad \varepsilon_{t} \sim \mathcal{N}\left(0, 1\right) \\
\end{equation}

\begin{equation}
\beta_{j, t} =\underline{\beta}_{j}+\rho\left(\beta_{j, t-1}-\underline{\beta}_{j}\right)+\underline{\delta} \eta_{j, t}, \quad \eta_{j, t} \sim \mathcal{N}\left(0, 1\right) \\
\end{equation}

\begin{equation}
\gamma_{j, t} =\underline{\gamma}_{j}+\rho\left(\gamma_{j, t-1}-\underline{\gamma}_{j}\right)+\underline{\delta} \nu_{j, t}, \quad \nu_{j, t} \sim \mathcal{N}\left(0, 1\right) \\
\end{equation}

\begin{equation}
\log \left(\sigma_{t}^{2}\right) =\underline{\sigma}^{2}+\phi\left(\log \left(\sigma_{t-1}^{2}\right)-\underline{\sigma}^{2}\right)+\underline{\xi} \zeta_{t}, \quad \zeta_{t} \sim \mathcal{N}\left(0, 1\right) \\
\end{equation}

\begin{equation}
\beta_{j, 0} =\underline{\beta}_{j}, \quad  \gamma_{j, 0} =\underline{\gamma}_{j}, \quad \log \left(\sigma_{0}^{2}\right)=\underline{\sigma}^{2}
\end{equation}

\noindent where the only changes will be related to the contemporaneous dependencies among series, i.e., for each subperiod all series will have different parents, depending on the ordering and respecting the triangular structure of (\ref{eq:triangular}). Our specification uses $\underline{\beta}_{j} \sim \mathcal{N}\left(0, 0.04\right)$,  $\underline{\gamma}_{j} \sim \mathcal{U}\left(-1, 1\right)$,  $\underline{\sigma}^{2} = 0.1$, $\rho = \phi = 0.99$, T = 600 and $\underline{\delta} = \underline{\xi} = T^{-2/3}$.

\begin{figure}[!h]
\begin{center}
	\includegraphics[width=17cm]{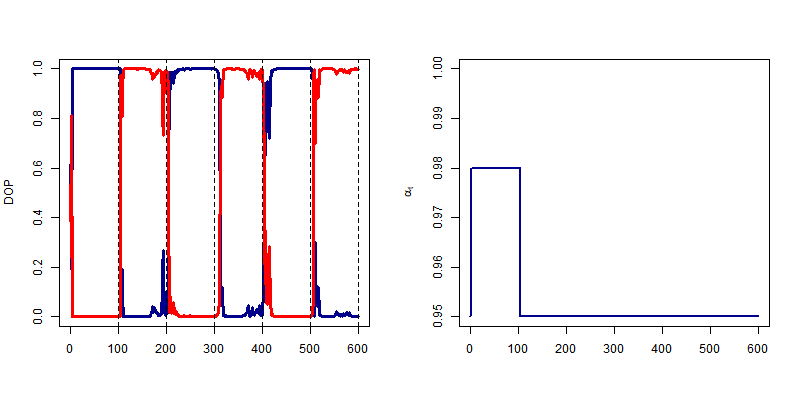} 
\caption{Left panel: Dynamic ordering Probabilities (DOP) for the two true simulated orderings over time. Vertical lines represent changes between the two true orderings. Right panel: Time-varying forgetting factor $\alpha_{t}.$
}
\label{fig: fig1}
\end{center}
\end{figure}

We apply our DOL approach to the generated data allowing the model to dynamically select the best forgetting factor $\alpha$ within the grid $\alpha \in \{0.95, 0.96, 0.97, 0.98, 0.99, 1 \}$ as explained in the previous section. The left panel of Figure \ref{fig: fig1} shows the dynamic posterior ordering probability (DOP) for the orderings contemplated on the simulations. The blue line is the DOP for the true ordering for the first, third and fifth subperiod while the red line is the DOP for the true ordering for the second, fourth and sixth subperiods. The vertical dotted lines are representing changes on the true ordering over time. Figure \ref{fig: fig1} makes clear how our approach is capturing the changes in the dynamic dependencies with great sucess. We can note that as soon as a new subperiod starts, the DOL method start to assign higher probabilities for the new true ordering and the probability of the previous one decreases very fast. The right panel of figure \ref{fig: fig1} shows the selected values for $\alpha$. After the first subperiod, our method is always selecting $\alpha = 0.95$, i.e, it is considering a lower forgetting factors for all subsequent periods. In fact, in this synthetic scenario where the correct ordering is continuously changing, this behavior is expected. Since lower $\alpha$ tends to discount more heavily past performance, it allows the model to switch faster to new dependency structures over time. In summary, this simulation study gives evidences that in a case where there are different ordering dependencies, our DOL method is quite able to dynamically learn from past mistakes, assigning the correct structure via DOP over time.

After introducing the DDNM framework and the DOL procedure, in the next sections we perform two different studies where we explore the main advantages of our econometric approach. In the first next section, we start by examining a portfolio allocation application and at Section (\ref{ref: Macro_section}) we analyze a macroeconomic forecasting exercise.

\section{Portfolio Allocation Problem}
\label{ref: Portfolio_section}

In this section we perform a dynamic asset allocation by combining the DDNM and DOL methods. The econometric application will be based on forecasts of a set of exchange rates and then we use the predictive information to sequentially rebalance the investor's portfolio. The motivation in proposing different methods to predict exchange rates is not new in the literature. The seminal work of \cite{meese1983empirical} brought evidences that structural models suffer to outperform a simple random walk. Since their work, the literature has been moving towards producing models that can generate better results in terms of out-of-sample accuracy or trading strategies. As summarized by \cite{rossi2013exchange}, each part of the literature use different predictors, models and approaches. They can differ in terms of the selected predictors, time-variation in parameters and the use of multivariate or univariate models. Just to name a few interesting studies on the applied econometrics for exchange rate predictability, we can refer to
\cite*{della2009economic}, \cite{della2012statistical}, \citeauthor*{byrne2016exchange} (
\citeyear{byrne2016exchange}, \citeyear{byrne2018sources}) on the univariate context and \cite*{beckmann2020exchange} in a multivariate application. 
 
Similar to \cite{della2009economic}, \cite{byrne2018sources} and \cite{beckmann2020exchange}, the dynamic portfolio allocation takes the perspective of an US investor who allocates her wealth between six foreign bonds and one domestic bond (US). At each period, each foreign bond yields a riskless return in the local currency and a risky return from the currency fluctuations in US dollars.

The investor takes two steps sequentially over time. The first is to use the econometric method to generate one-month ahead forecasting. In the second step, the investor dynamically rebalances the portfolio by finding new optimal portfolio weights. To perform the portfolio optimization, the investor will use the vector of predicted mean exchange rate returns and the predicted covariance matrix. Using this setup we are able to assess the economic value of exchange rate predictability from different methods within a dynamic mean-variance framework, implementing
a maximum expected return strategy subject to a conditional volatility target.

Following \cite{della2009economic} and \cite{byrne2018sources}, let $\boldsymbol{r}_{t+1}$ be the $m \times 1$ vector of risky asset returns, $\boldsymbol{\mu}_{t+1|t} = E_{t} [ \boldsymbol{r}_{t+1} ]$ and $\boldsymbol{\Sigma}_{t+1|t}$ the $m \times m$ conditional covariance matrix of $\boldsymbol{r}_{t+1}$. At each period of time, the investor solves the following problem:
\begin{equation}
\max _{\boldsymbol{w}_{t}}\left\{\boldsymbol{\mu}_{p, t+1}=\boldsymbol{w}_{t}^{\prime} \boldsymbol{\mu}_{t+1 \mid t}+\left(1-\boldsymbol{w}_{t}^{\prime} \boldsymbol{\iota} \right) r_{f}\right\} \ \ \ \mbox{such that} \ \ \sigma_p^{2*}=\boldsymbol{w}_{t}^{\prime}\boldsymbol{\Sigma}_{t+1|t}\boldsymbol{w}_{t}
\end{equation}
where $\boldsymbol{\iota}$ is a vector of ones, $\mu_{p, t+1}$ is the conditional expected portfolio return, $\sigma_{p}^{*}$ is the volatility target, $\boldsymbol{w}_{t}$ is the $m \times 1$ vector of new portfolio weights and $r_{f}$ is the return of the riskless asset.  In our study we consider an annualized volatility target of $\sigma_{p} = 10\%$.  \cite{della2009economic} show that the solution to this problem implies the following weights to the risky bonds:
$$
\boldsymbol{w}_{t}=\frac{\sigma_{p}^{*}}{\sqrt{C_{t}}} \boldsymbol{\Sigma}_{t+1 \mid t}^{-1}\left(\boldsymbol{\mu}_{t+1 \mid t}-\boldsymbol{\iota} r_{f}\right),
$$
where
$C_{t}=\left(\boldsymbol{\mu}_{t+1 \mid t}-\boldsymbol{\iota} r_{f}\right)^{\prime} \boldsymbol{\Sigma}_{t+1 \mid t}^{-1}\left(\boldsymbol{\mu}_{t+1 \mid t}-\boldsymbol{\iota} r_{f}\right)$. 
The gross return of the investor's portfolio is computed as 
$$
R_{p, t+1}=1+r_{p, t+1}=1+\left(1-\boldsymbol{w}_{t}^{\prime} \boldsymbol{\iota} \right) r_{f}+\boldsymbol{w}_{t}^{\prime} \boldsymbol{r}_{t+1}.
$$
Differently from \cite{della2012statistical} and \cite{byrne2018sources}, where the authors do not model
the conditional covariance matrix of exchange rate returns and just replace  $\Sigma_{t+1 \mid t}$ by the unconditional covariance matrix, we use instead the predicted covariance matrix estimated from our econometric model.

\subsection{Model Assessment}
\label{model_assessment}

We briefly explain the main criteria used to compare different approaches in terms of out-of-sample forecasting accuracy and economic performance. In general, it is common in the econometric literature to consider point and density forecasting metrics. However, in this portfolio allocation section, the investor is not just concerned about forecasting accuracy, but also how this accuracy is translated to better portfolio performance and utility improvements for a mean-variance investor. Therefore, after introducing the main statistical evaluation measures, it is crucial to explain the main economic criteria used to evaluate the outputs of our econometric method.

\textbf{Statistical Evaluation:} We compare point and density forecast metrics, where point forecast accuracy assessment will be given by the Mean Square Forecast Error (MSFE). Let $MSFE^{L}_{j}$ represents the MSFE of currency $j$ produced by a specific model $L$. We compare point forecasts of different models compared to a benchmark model by

\begin{equation}
MSFE^{L}= \frac{\sum_{j=1}^{m} MSFE_{j}^{L}}{\sum_{j=1}^{m} MSFE_{j}^{Bmk}}
\end{equation}

\noindent where $m$ is the total number of series to be forecasted and $MSFE_{j}^{Bmk}$ is the MSFE of series $j$ produced by the benchmark model. Motivated by the evidence of stronger performance on random-walk models in the exchange rate predictability literature, in this portfolio allocation study we consider a simple multivariate Random-Walk as a benchmark model, where the error covariance matrix is assumed to follow an inverted Wishart distribution (W-RW)\footnote{We have considered in our study a driftless Random Walk model with time-varying covariance matrix ($\kappa = 0.96$)}.

In terms of density forecast we use the Log Predictive Density Ratio ($LPDR$), following the recent Bayesian econometric literature (\citealp{mcalinn2019dynamic},  \citealp{mcalinn2020multivariate}, \citealp{nakajima2013bayesian} and \citealp{koop2013large}).  In this metric, we use the predictive density for $\boldsymbol{y_{t}}$ (given all data available until $t-1$ evaluated at the actual outcome, $p(\boldsymbol{y_{t}}| \boldsymbol{D_{t-1}}).$ We opted by this criteria because our interest
here is not just making point forecasting, but also to generate better predictions about the whole predictive distribution. In out first application, the mean and the variance of the predictive density will be essential to build portfolios. Therefore, a density
forecast criteria suits much better to our interest. Additionally, it aligns with the work of
\cite{cenesizoglu2012return} who shows evidence of agreements between density
forecast and economic performance.

The density forecast criteria ($LPDR$) is defined as the ratio between the sum of the log-predictive density of model $L$ and the sum of the log-predictive density of the benchmark model :
\begin{equation}
\operatorname{LPDR}_{L}= \sum^{T}_{t=1} \log\left\{
\frac{p_{L}\left(\boldsymbol{y_{t+1}} | \boldsymbol{y_{t}}\right)}
{p_{Bmk}\left(\boldsymbol{y_{t+1}} | \boldsymbol{y_{t}}\right)}
\right\}.
\end{equation}
The LPDR provides a statistical assessment of relative accuracy that extends traditional
Bayes’ Factor. Therefore, whenever $LPDR_{L} >
0$, it means that model $L$ is statistically outperforming the benchmark.

\textbf{Economic Evaluation:} In order to evaluate the economic performance of our DOL method for exchange rate predictability and portfolio allocation, we use a standard mean-variance measure. Following \cite*{fleming2001economic}, we compute ex-post average utility for a mean-variance investor with a quadratic utility.  As in \cite{fleming2001economic}, \cite{della2009economic} and \cite{beckmann2020exchange} we can calculate the performance fee that an investor will be willing to pay to switch from the tradicional Wishart-Random Walk (W-RW) model to the DOL approach. The performance fee is computed by equating the average utility of the W-RW portfolio with the average utility of the DOL portfolio (or any alternative portfolio), considering the latter with a fee $\Phi$:

$$\sum_{t=0}^{T-1}\{(R_{p, t+1}^{DOL}-\Phi)-\frac{\gamma}{2(1+\gamma)}(R_{p, t+1}^{DOL}-\Phi)^{2}\} =  \sum_{t=0}^{T-1}\{R_{p, t+1}^{RW}-\frac{\gamma}{2(1+\gamma)}(R_{p, t+1}^{RW})^{2}\}$$

\vspace{4mm}

\noindent where $\gamma$ is the investor's degree of relative risk aversion (RRA), $R_{p, t+1}^{DOL}$ is the gross return from the DOL portfolio and $R_{p, t+1}^{RW}$ is the gross return from the W-RW portfolio. As in \cite{beckmann2020exchange}, we set $\gamma = 2$ in our main results. In Appendix C we show additional results using different  $\gamma$'s and relative performances are still robust.
In this way, we can interpret $\Phi$ as the maximum annualized performance fee an investor is willing to pay to switch from a Wishart multivariate Random Walk model  to the Dynamic Ordering Learning (DOL) approach.

In our following results below, we also show Sharpe Ratios (SR) as an additional economic performance measure. It is the most commonly used measure in the financial literature and among practitioners. This measure is the average excess return of the portfolio divided by the standard deviation of the portfolio returns. 

All economic measures displayed in Section \ref{sec: Econ_Empirical} are already net of transaction costs (TC). Following \cite{marquering2004economic}, we deduct the
 transaction cost from the portfolio return ex-post. As argued by \cite{della2012statistical}, it is a reasonable simplification that maintains the tractability of the analysis. Differently from \cite{della2012statistical} and \cite{beckmann2020exchange}, that have set the transaction cost at $8$bps, we decide to use a slightly more conservative transaction cost of $10$bps.

\subsection{Empirical Results}
\label{sec: Econ_Empirical}

As predictors, we propose to sequentially select one of 12 different measures of time series momentum, each one with a specific look-back period, ranging from 1 to 12 months. The time series momentum predictor is a measure of continuation (or trend) in returns. Its ability to predict returns is well documented in the financial literature (\citealp{jegadeesh1993returns}, \citealp{moskowitz2012time} and \citealp{levy2021time}) and it can be defined as the accumulated returns from the previous $l$ months, where $l$ is the size of the look-back period\footnote{Specifically, we can define a momentum measure of look-back period $l$ as: $MOM_{t}^{l} = \frac{P_{t}}{P_{t-l}} -1 $}. Additionally, each univariate model will have a different pair of discount factors $\delta$ and $\kappa$, with the following possible value choices for each one: $\delta \in$ \{0.99, 1\} and $\kappa \in $  \{0.96, 1\}. Hence, the investor is able to sequentially learn not just about how far she needs to look into the past to infer about the best trend to predict returns but also is able to learn if coefficients and volatilities are constant or if they are time-varying, since discount factors lower than one induce variation in parameters and discount factors equal to one induce constant parameters\footnote{See Appendix A for more details.}. Therefore, when we refer to our DOA or DOS approaches we are meaning a model in which variation in coefficients are induced but constant parameters are also allowed if in some periods it is empirically wanted.

The dataset consists of a set of six most traded currencies: the Australian dollar (AUD), the Canadian dollar (CAD), the Euro (EUR), the Japanese yen (JPY), the Swiss franc (SWF), the Great Britain pound (GBP) and the US dollar (USD) and by the one-month LIBORs for the respective countries. All currencies are expreesed in terms of the US dollar and are end-of-month exchange rates, computed as discrete returns. All the data in our application study here was taken from the work of \cite{beckmann2020exchange}\footnote{Available at \url{https://sites.google.com/site/dimitriskorobilis/matlab/fx_tvp} }. The sample runs from 1986:01 until 2016:12 and we use the first ten years of data as training period and the last twenty years as statistical and economic out-of-sample evaluation period. 


\begin{figure}[!h]
\begin{center}
	\includegraphics[width=17cm]{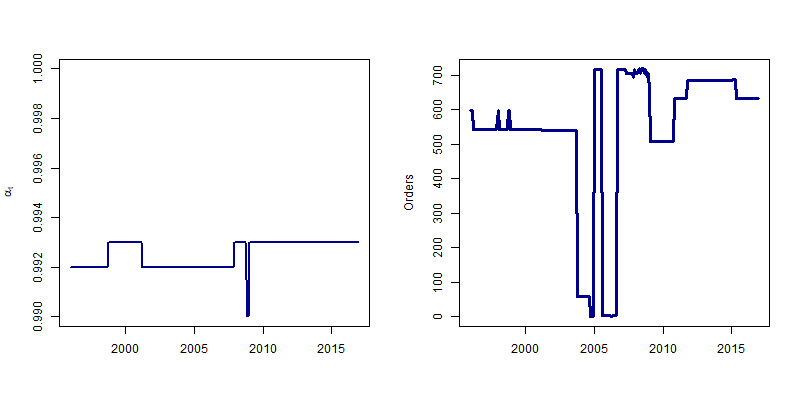} 
\caption{Time-Varying Forgetting Factor $\alpha_{t}$ (Left panel)  and ordering Selection (Right panel). The ordering Selection panel shows specific orderings with the highest ordering probability for each period of time.}
\label{fig: figura1}
\end{center}
\end{figure}

 We use our DOL approaches to forecast the mean and covariance matrix of exchange rate returns. This study will compare statistically and economically the use of DOA and DOS with a mean-variance investor that uses a simple multivariate Random-Walk, where the error covariance matrix is assumed to follow an inverted Wishart distribution (W-RW). Since our multivariate model contains six currencies, we have a total of 6! = 720 possible ordering permutations. Hence, we also show results compared to the case where the investor believes in models that uses fixed orderings for all periods of time. We show evidences that the importance of orderings are in fact dynamic, meaning that learning the chances in the contemporaneous relations among currencies improves statistical measures and portfolio performance.



Within each ordering structure, the investor can also learn about the look-back period for a time-series momentum strategy and different degrees of variability in parameters over time. Since the environment of the economy induces changes in the behavior of returns, with periods of faster or slower switchings among orderings, we allow to select among eleven values for the forgetting factor $\alpha$ over time\footnote{ We choose $\alpha \in \{0.99, 0.991, 0.992, 0.993, 0.994, 0.995, 0.996, 0.997, 0.998, 0.999, 1 \}$}. The left panel of Figure \ref{fig: figura1} reports the selected $\alpha$ across time. We can notice that in all periods of time the best forgetting factors were below 1, which means that indeed we have a dynamic switching behavior among orderings over time, with periods of faster and slower changes. For most of the time, it fluctuates between $\alpha = 0.992$ and $\alpha = 0.993$, with a exception at the end of 2008, where it reached 0.990 for some few months.

In terms of the orderings that received the highest probability over time, the right panel of Figure \ref{fig: figura1} shows a continuous changing. It is interesting to notice that there is no best ordering for all periods of time or for long periods, in the sense that an ordering that performed well at the end of the 90s is not the same as the best ordering during the 2005 or during the Great Recession. In fact, there is a high instability around the orderings during the period of 2007-2008, where the best ordering is continuously changing almost at the monthly frequency.



This results are in favor with our idea of ordering uncertainty and different contemporaneous relations among currencies. Since the ordering with the highest ordering probability is switching over time, with periods of stronger instability and periods of calm behaviors, we have evidences that those orderings are performing differently over time and taking into account uncertainty around them is crucial for statistical analysis and decision-making, as we show at the next subsections.   

\vspace{4mm}

\subsubsection{Statistical Performance}
\label{sec: Statistical Evaluation}

\begin{figure}[h!]
\begin{center}
	\includegraphics[width=17cm]{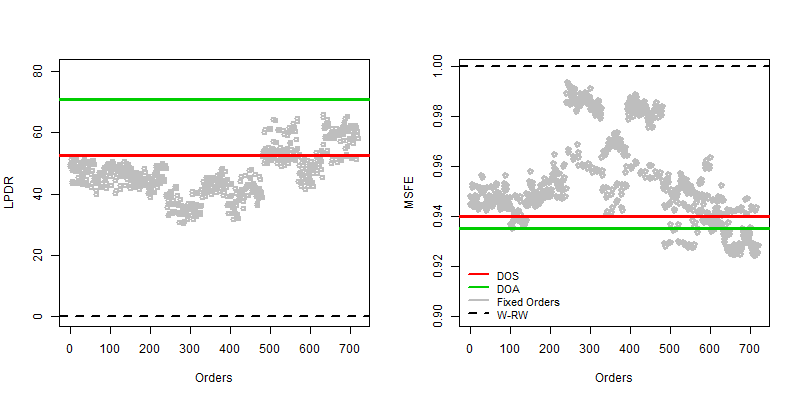} 
\caption{Statistical performance relative do the Wishart-Random-Walk model. i) Left panel: Log Predictive Density Ratio (LPDR); ii) Right panel: Mean Square Forecast Error (MSFE)}
\label{fig: figura2}
\end{center}
\end{figure}

We present point and density forecast evaluation for the period of 1996:1 through 2016:12. Both measures will assess how our DOL approach performs in terms of one-month ahead forecasting out-of-sample. As previously mentioned, we use mean square forecast error (MSFE) as a measure of point forecast and, for density forecast, the predictive likelihood. The latter is popular in the Bayesian literature and captures how the whole predictive distribution performs to forecast and not just a single value opposed to the MSFE. The right panel of Figure \ref{fig: figura2} shows the MSFE performance of both of our DOL approaches, the DOA (green line) and DOS (red line).  The gray points are the point forecast performance of all 720 possible fixed orderings structures that the econometrician/investor has available. The MSFE is in relation to the Wishart-Random Walk (W-RW) model and numbers lower than one means that the specific model is outperforming the W-RW model. As it is clear, all different specifications outperform the Random Walk model. Both the DOA and DOS approaches are performing better than the huge majority of the fixed orderings. Also, there are great differences among the performance between the fixed ordering possibilities, meaning that relying on the use of a random fixed ordering for all periods of time can make huge differences at final outcomes. 

\begin{figure}[!h]
\begin{center}
	\includegraphics[width=10cm]{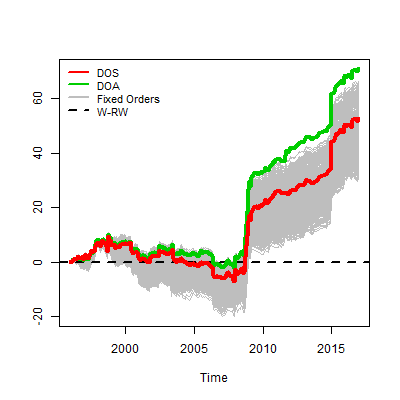} 
\caption{Accumulated Log Predictive Likelihood relative to the Wishart-Random Walk Model}
\label{fig: figura3}
\end{center}
\end{figure}


In relation to density forecast performance, the left panel of Figure \ref{fig: figura2} shows how the DOA outperforms not just the Random Walk model, but all 720 fixed orderings over time.\footnote{Numbers above zero represents that the specific model is performing better than the benchmark (W-RW) in terms of density forecast.} Hence, when the investor sequentially learn about the importance of each ordering and considers the fact that one ordering can improve or get worse in matters of quarters or even months, the statistical performance increases a lot. Also note that the DOS performs better than the vast majority of  the fixed orderings.

In order to understand how the density performance evolve over time, we plot at Figure (\ref{fig: figura3}) the accumulated predictive likelihood. This metric is useful to visualize how different models are accumulating statistical gains or losses compared to the benchmark over time. We can note that the DOA approach is the best model among all fixed orderings for the whole out-of-sample evaluation period and accumulates significant gains in relation to the benchmark, specially after the Great Recession.

\subsubsection{Portfolio Performance}
\label{sec: Economic Evaluation}


The previous subsection provided evidences that the DOL approach is improving statistical performance and capturing ordering change.  But nothing was said yet about portfolio and utility improvements. As we have previously highlighted before, we design a portfolio allocation study in which an US investor optimally allocates her wealth among six foreing bonds and the US bond, receiving not just the riskless return from those bonds but also the risky currency fluctuations. We compute annualized Sharpe Ratios (SR) for the investor who uses different econometric models to generate predicted mean and covariances among currencies to rebalance her portfolio each time period. We also show the Annualized Management Fee ($\Phi$) that the investor will be willing to pay to switch from the W-RW model to each one of the methods (DOA, DOS or fixed orderings).

\begin{figure}[!h]
\begin{center}
	\includegraphics[width=17cm]{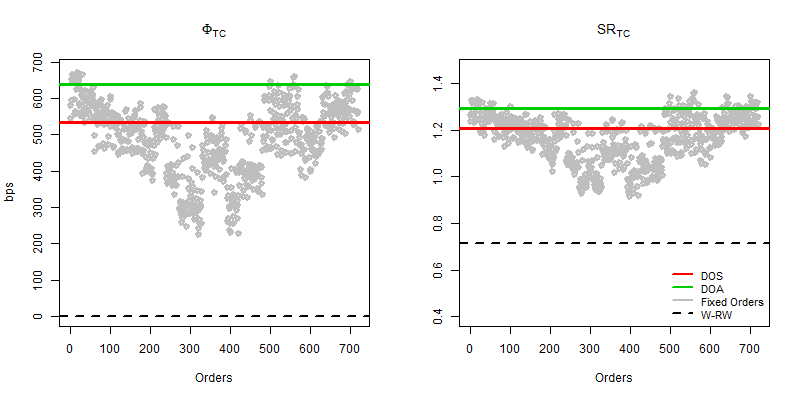} 
\caption{Economic performance relative do the Wishart-Random-Walk model. i) Left panel: Annualized Management Fees ($\Phi$); ii) Right panel: Sharp Ratios. All results are already net of transaction costs (TC = 10 bps).}
\label{fig: figura4}
\end{center}
\end{figure}


The right panel of Figure (\ref{fig: figura4}) shows the SR for different strategies. The results are already net of transaction costs. The Random Walk strategy generates an annualized Sharpe Ratio of 0.71 for the out-of-sample evaluation period. It is not a bad performance, but as the figure makes clear, any fixed ordering generate much better average return adjusted for risk for the investor. The green line shows how our DOA strategy dramatically improves SR for the investor. An investor using our DOA approach would generate a portfolio with SR of 1.30 for the period, a number greater than around 92\% of all fixed orderings over time. The DOS strategy also performed well, with a SR equal to 1.21, also much higher than the W-RW approach. Again, although all fixed orderings have produced good portfolio results, there are considerable differences in final performance among them, meaning that ordering uncertainty plays an important role on final outcomes.

It is important to highlight that although Sharpe Ratio is a popular measure among practitioners, it tends to overestimate risk for dynamic portfolios (\citealp{marquering2004economic} and \citealp{beckmann2020exchange}). It motivates the use of a more robust measure of economic performance, considering explicitly the risk aversion and a utility function for the investor (as explained in Section \ref{model_assessment}). In terms of economic utility, the left panel of Figure (\ref{fig: figura4}) shows the annualized management fee (net of transaction costs) that a mean-variance investor would pay to switch from the W-RW to the proposed methods. This figure makes clear the strong performance of the DOA strategy. In fact, a mean-variance investor will be willing to pay the considerable fee of 638.5 basis points to migrate to the DOA strategy. The DOA requires a fee that is higher than around 98\% of the fees for all fixed orderings. As we have noted for the statistical measures and SR, the investor that considers a fixed ordering over time will give up the opportunity to learn the time-varying contemporaneous dependencies among currencies and will be subject to a large variance on possible final outcomes.

\begin{figure}[!h]
\begin{center}
	\includegraphics[width=17cm]{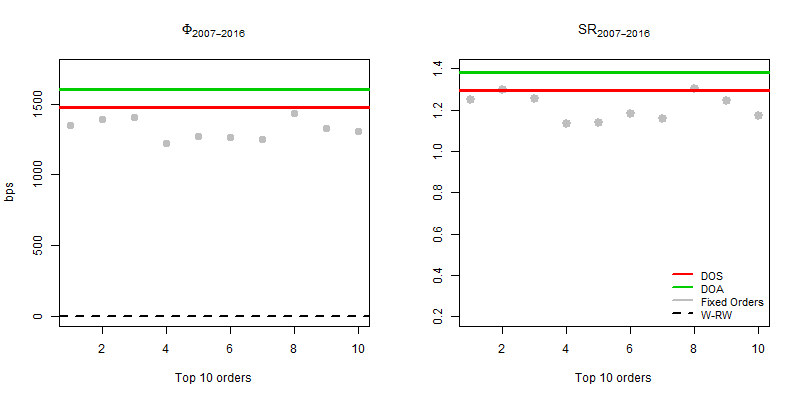} 
\caption{Economic Performance 2007-2016: DOA and DOS against top 10 orderings at the end of 2006.}
\label{fig: figura5}
\end{center}
\end{figure}

Although one may argue that  the DOA and DOS strategies generate strong portfolio performances, it still remains some few fixed orderings that have shown even higher performance and, possibly, it would be worth to consider this small set of fixed orderings as forecasting model candidates, using them to generate predictions as inputs in the portfolio allocation. However, we highlight here that the investor was not aware of the performance of those fixed orderings in advance. When performing an asset allocation considering the best ordering structures, the investor should consider those orderings that have performed better at the time of the decision. Hence, we investigate what an investor would have done in terms of portfolio allocation if she had considered the top fixed orderings at the end of 2006 and then allocated her wealth using them compared to our DOL approach. We consider an investor who observes all the data available until December of 2006 and computed the management fee that all fixed orderings have generated compared to the W-RW. Figure (\ref{fig: figura5}) shows the out-of-sample performance between 2007 and 2016 that the investor would get if she just believed on the superiority of the top 10 fixed orderings in terms of management fee at that time and used those 10 models to allocate her wealth. What we can confirm from this figure is that when the investor gives up the opportunity to learn about changes in importance among different orderings and avoids the dynamic uncertainty on the dependencies of each currency over time, the final economic performance is harmed. The investor who instead considers the fact that there is strong uncertainty about the correct ordering structure and it is continuously changing depending on the environment of the economy finishes the out-of-sample period with not just a higher Sharpe Ratio, but a much higher utility gain. 

Therefore, we argue that is quite difficult to antecipate what is a good ordering in advance and even a good guess can lead to a suboptimal future performance. Our approach is able to recognize which orderings are starting to perform better or worse in a dynamic fashion and then attribute higher or lower probabilities to them. 

Finally, in Appendix C we also show the results for different model settings in terms of variation in parameters and volatilities compared to the W-RW. Table 1 and Figure (\ref{fig: figura_appendix1}) make clear the great statistical superiority of ordering averaging and time-varying coefficients (TVP and SV) for point and density forecast. The statistical gains are in general even higher when compared to models with constant volatilities (CV). Table 2 shows the great economic performance of different model settings, specially when compared to the Wishart Random-Walk with constant volatilities (W-RW-CV). We report management fees for different levels of risk aversions ($\gamma$). Interestingly, both DOA and DOS with CV generate slightly higher utility gains compared to their SV counterparts for the smallest risk aversion. However, as soon as the risk aversion increases, the SV setting appears with much stronger performance. In terms of SR, combining both TVP and SV also have produced higher out-of-sample portfolio performance.

\section{Macroeconomic Forecasting}
\label{ref: Macro_section}

Vector Autoregressive (VAR) models are commonly applied in the macroeconomic literature and used in Central Banks and financial institutions in many different contexts. VARs are known to be a powerful tool to predict the future movements of the economy and for monetary policy evaluation (\citealp{sims1980macroeconomics}, \citealp{litterman1986forecasting},  \citealp{primiceri2005time}, \citealp{clark2010averaging} and \citealp{koop2013large}, \citealp{kastner2020sparse}).

The recent VAR literature has recognized the advantages of considering time-varying parameters and volatilities when building forecasting models. Inspired by the Cholesky-style behind the work of \cite{primiceri2005time} and \cite{del2015time}, we are motivated to explore the ability of our approach to deal with the problem of ordering uncertainty in a macroeconomic context. Since the macroeconomy is continuously adapting to new environments and different sources of breaks, such as wars, global  crisis and pandemics,  VAR models are strongly susceptible to instabilities, as highlighted for instance in \cite{cogley2005drifts} and \cite{clark2010averaging}. We argue here that when the main goal of the econometrician is to produce sequential forecasts, those instabilities can induce different sources of dependencies among economic variables. However, since the Cholesky-style framework is tied to the ordering structure, the out-of-sample forecasting results can be seriously harmed from the static behavior of economic series dependencies, which can change rapidly from year to year or just in few months. 

It is important to highlight that, when we allow our model to learn and explore different series dependencies over time, our interest here is not on identification assumptions or challenging economic theories behind those dependencies, but instead focus on improving out-of-sample forecasting accuracy. 

Therefore, we will follow a similar DDNM structure made by \cite{zhao2016dynamic}, where now the predictors will be composed by the time series lagged values, building on the format of VARs with time-varying parameters and stochastic volatilities (TVP-VAR-SV). Similar to \cite{primiceri2005time} and \cite{del2015time}, we will focus on a VAR model with three important US macroeconomic variables: inflation, unemployment and interest rates.


\subsection{Empirical Results}

As described before, we use the DDNM framework of \cite{zhao2016dynamic} combined with our DOL approach to build TVP-VAR-SV models that are able to sequentially learn the contemporaneous relations among inflation, unemployment and interest rates via dynamic ordering probabilities. These macroeconomic series were also considered in the small-scale VAR of \cite{primiceri2005time}. We use quarterly data for the US economy from 1953Q1 to 2015Q2. We left the first 150 quarters (until 1990Q2) as training period and perform an out-of-sample evaluation for the next 100 quarters (from 1990Q3 to 2015Q2). Inflation is measured as the year-over-year log growth rate of the GDP price index. Unemployment rate is referred to all workers over 16 years and interest rate is the yield on 3-month Treasury bills\footnote{The GDP price index can be obtained from the Federal Reserve Bank of Philadelphia (\url{https://www.philadelphiafed.org/surveys-and-data/real-time-data-research}). Unemployment and interest rate can be downloaded from the Federal Reserve Bank of St. Louis (\url{https://fred.stlouisfed.org/}). The data is also easily accessible on the R package \textbf{bvarsv} (\citealp{krueger2015bvarsv}).}

\begin{figure}[!h]
\begin{center}
	\includegraphics[width=17cm]{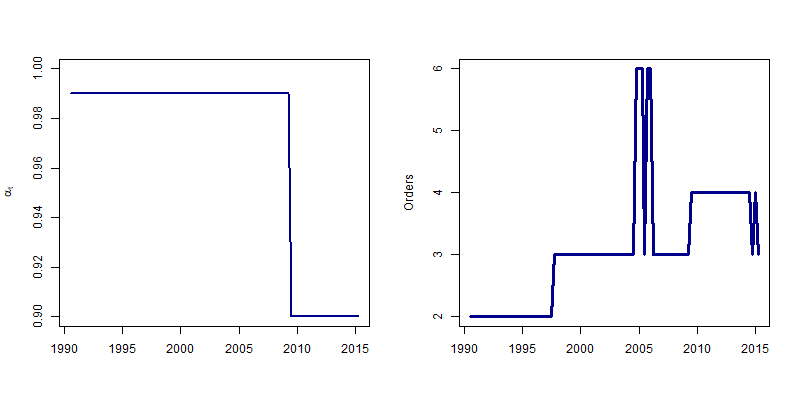} 
\caption{Time-Varying Forgetting Factor $\alpha_{t}$ (Left panel)  and ordering Selection (Right panel). The ordering Selection panel shows specific orderings with the highest ordering probability for each period of time.}
\label{fig: figura_macro1}
\end{center}
\end{figure}


We set our TVP-VAR-SV model to sequentially learn the use of two lags of all dependent variables. Each equation can adjust to the use of each lag predictor to enter or not in the model for each period of time. Since we use quarterly data, it is common practice to induce a higher discount in information, since few observations are able to contain long periods of time and just in few quarters of data the environment of the economy can dramatically change. Therefore, we consider the range of values for the discount factors $\delta$ and $\kappa$ $\in$ $\{0.95, 0.99, 1\}$. As we did in the portfolio allocation problem, we still let our approach to learn the degree of variation in coefficients, switching from higher and lower degrees of variation to a constant coefficient if it is empirically wanted. \footnote{In term of the forgetting factor $\alpha$, we also allow for higher decay in model probabilities, selecting among $\alpha \in \{0.9, 0.91, 0.92, 0.93, 0.94, 0.95, 0.96, 0.97, 0.98, 0.99, 1 \}$.}


This small-scale VAR has just three economic dependent variables, implying the existence of $3! = 6 $ possible ordering structures to consider. \cite{primiceri2005time} considers a standard identification assumption where monetary policy actions affect inflation and unemployment with at least one period of lag, which means that interest rate is placed last in the ordering structure. In his work, the ordering of inflation and unemployment were arbitrarily selected to enter at first and second, respectively. We use the ordering setting of \cite{primiceri2005time} as a benchmark in our study, comparing how our DOL approach and using fixed orderings over time would perform in relation to this traditional assumption. Note that the specific ordering of $ \boldsymbol{y_{t}} = \left[Inflation_{t} , \quad Unemployment_{t} , \quad Interest Rate_{t} \right]'$  is one of the six possible fixed orderings over time. Therefore, in our plots we show how DOL and others fixed orderings perform in relation to this benchmark. Again, we argue here that the main goal of the study is to check for out-of-sample predictability instead of conjecture about identification assumptions and macroeconomic theory.

The left panel of Figure (\ref{fig: figura_macro1}) shows the time-variability of the forgetting factor $\alpha$. During all the evaluation period, it remains lower than one, meaning that a higher discount on predictive densities are induced. After the Great Financial Crisis, the model selected a even lower $\alpha$, which means that orderings that performed well in the very recent past are preferred to orderings that performed well in the past. This behavior can be strongly related to changes in the economic behavior. Finally, the right panel of Figure (\ref{fig: figura_macro1}) shows the series orderings that received the highest order probability for each period of time. Note that there is no single ordering that dominates others for the evaluation period. Interestingly, for the whole out-of-sample period, the standard ordering used in \citealp{primiceri2005time} (the ordering number one in the right panel) was never selected as the best ordering structure and preferred to the others.


\subsubsection{Statistical Performance}
\label{sec: Statistical Evaluation}

Unlike the portfolio allocation problem, in the context of macroeconomic forecasts, we will focus only on measures of statistical accuracy: MSFE and LPDR. As a benchmark (in black), we select a fixed ordering within our DDNM model structure such that this ordering is the same standard ordering used in \cite{primiceri2005time}, where both inflation and unemployment are affected by monetary policy after at least one lag of time. Hence, although the benchmark allows to learn different lag predictors and variation in coefficients, it does not consider series ordering switching and learning.  We also present the performance of the remaining 5 fixed orderings (gray dots). Again, the green line represents the statistical performance of the DOA approach and in red the DOS performance.

\begin{figure}[h!]
\begin{center}
	\includegraphics[width=17cm]{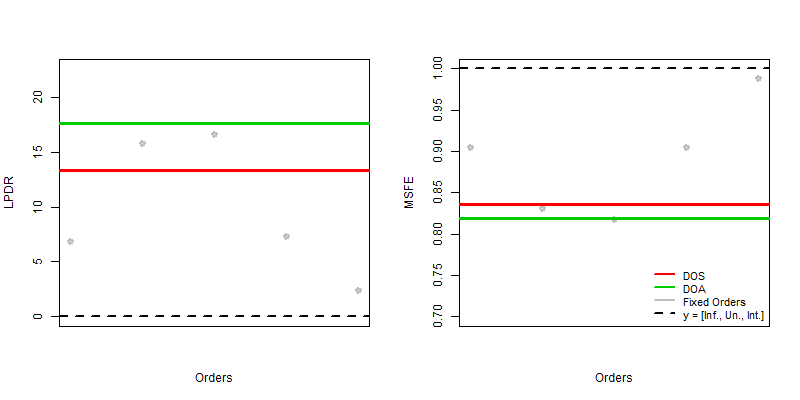} 
\caption{Relative statistical performance relative to the benchmark. i) Left panel: Log Predictive Density Ratio (LPDR); ii) Right panel: Mean Square Forecast Error (MSFE)}
\label{fig: figura_macro2}
\end{center}
\end{figure}


In terms of point forecast, the right panel of Figure (\ref{fig: figura_macro2}) shows the great superiority of DOA and DOS approaches in relation to the benchmark, representing the important out-of-sample accuracy  improvement for the econometrician that learns sequentially from the data differences in the dynamic contemporaneous dependencies among macroeconomic variables over time. There are two fixed orderings that performed quite similar to DOS and DOA. Those series orderings are, interestingly, considering inflation at the bottom of the ordering structure instead of the top. Hence, for these two orderings, inflation is being affected by monetary policy and unemployment contemporaneously. Also, monetary policy is responding to inflation with at least one lag of time.

In relation to density forecasting, the results are similar. DOA performed better than all fixed orderings and the benchmark. Also note that any fixed ordering and the DOS approach outperform the standard benchmark ordering structure. Again, those series orderings with inflation at the last position in the $\boldsymbol{y_{t}}$ vector have showed great improvements compared to the benchmark.


\begin{figure}[!h]
\begin{center}
	\includegraphics[width=10cm]{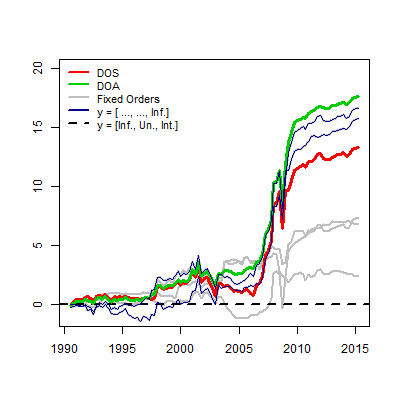} 
\caption{Accumulated Log Predictive Likelihood relative to the benchmark.}
\label{fig: figura_macro3}
\end{center}
\end{figure}

The relevance of considering inflation at the end of the ordering structure is highlighted in Figure (\ref{fig: figura_macro3}). It presents the accumulated log predictive density, so it is a measure of how an ordering structure is accumulating density forecast gains over time compared to the benchmark. In blue we emphasize those orderings that use inflation at the end of the $\boldsymbol{y_{t}}$ vector. Before 2007-2008, none of the orderings were easily seen as a superior. However, at the end of 2007 and beggining of 2008, the stastistical performance of those orderings in blue started to abruptly grow. This improvement growth lasted until around 2011 and, since then, they still maintain the accuracy gains obtained before. It seems that for those stressed periods, an ordering structure that considers a monetary policy contemporaneously independent of inflation predicted much better the future movements of the economy.

The great advantage of our DOL approach is that, as soon as an ordering structure captured some new kind of information in the economic environment, it starts to attribute more probabilities for those orderings. This framework allows the DOA to accumulates higher density forecast improvements than all other orderings.

Finally, we also show in Appendix C some additional robustness results for the macroeconomic application with different discount factors. At Table 3 and Figure (\ref{fig: figura_appendix2}), we can note the great statistical improvement of the DOA approach with time-varying parameters and volatilities compared to models considering only constant parameters. The statistical gains are even higher when compared to models with constant volatilities. The results are in line with the empirical evidences on time-varying volatilities patterns on economic series. Interestingly, ordering selection performed slightly better than the ordering averaging approach for some specific model settings. Table 3 shows evidences that models combining both TVP and SV deliver stronger out-of-sample forecasting performance.

\section{Conclusion}
\label{sec: Conclusion}

Since the recent and growing literature on multivariate forecast, where the popular Cholesky-style have been adopted as a flexible method to decouple a multivariate model into a set of univariate DLMs, little has been discussed about the differences in final decisions when considering different ordering structures. The main goal of our work is to solve this ordering uncertainty in an online fashion. We extend the class of Dynamic Dependency Network Models of \cite{zhao2016dynamic} and propose the Dynamic Ordering Learning approach, a very fast and flexible method to deal with the uncertainty around the contemporaneous relations among dependent variables. We perform a dynamic asset allocation study where the investor is uncertain about the contemporaneus relations among different currencies and we show that the Dynamic Ordering Learning approach generated not just significant statistical improvements, but also great economic gains for the investor. The results show that the mean-variance investor will be willing to pay a considerable annualized management fee to switch from the traditional Wishart Random Walk model to the DOL approach. Additionally, the DOA approach performs much better than the huge majority of models with fixed orderings. 

As a second application, we use a VAR structure within our DOL approach to forecast inflation, unemployment and interest rates. We show evidences that DOL is able to adapt to changes in the environment of the economy, giving higher probabilities for those orderings that have performed better in the recent past. We provide evidence that during the Great Financial Crisis, our approach detected great improvements when changing the dynamic dependencies among economic variables, incorporating this new information. We show that the DOL was able to substantially increase both point and density forecast accuracy compared to a standard orderingstructure commonly used in the macroeconomic literature. 

In summary, we found evidences that taking into account different contemporaneous relations among variables over time improves statistical models and final decisions, since the environment of the economy is continuously changing and the dependencies of variables are switching over time.  We highlight here that our framework can be expanded to a broader perspective, being applied not just on portfolio allocations or macroeconomic forecasting, but in any field where the researcher is faced with the problem of multivariate sequential forecasts.


\clearpage

\appendix
\setcounter{secnumdepth}{0}

\section{Appendix A: Filtering and Forecasting}
\label{appendix A}

We give details about the evolution and updating steps for the set of m univariate DLMs, following \cite{zhao2016dynamic} and similar to \cite{fisher2020optimal}. 

\noindent \textbf{Posterior at $t-1$}:  At time $t-1$ and for each series j, we define the initial states for $\theta_{j t-1} $ and volatility $\sigma_{j t-1}$ as:

\begin{equation}
\theta_{j, t-1}, \sigma_{j t-1}^{-1} \mid \mathcal{D}_{t-1} \sim \mathcal{NG}\left(\boldsymbol{m}_{j, t-1}, \boldsymbol{C}_{j, t-1}, n_{j, t-1}, n_{j, t-1} s_{j, t-1}\right)
\end{equation}

\vspace{5mm}

\noindent Equation (16) is the joint posterior distribution of model parameters at time $t-1$, known as a Normal-Gamma distribution. Hence, given the initial states, posteriors at $t-1$ evolve to priors at $t$ via the evolution equations:

$$
 \boldsymbol{\theta}_{j, t}=\boldsymbol{\theta}_{j, t-1}+\boldsymbol{\omega}_{j, t}, \quad \text { where } \quad \boldsymbol{\omega}_{j, t} \sim N\left(0, \mathbf{W}_{j, t} /\left(s_{j, t-1} \sigma_{j t}^{-1}\right)\right) 
 $$
 
 $$
 \sigma_{j t}^{-1}  = \sigma_{j t-1}^{-1}  \eta_{j, t} / \kappa_{j}, \quad \text { where } \quad \eta_{j, t} \sim \operatorname{Be}\left(\kappa_{j} n_{j, t-1} / 2,\left(1-\kappa_{j}\right) n_{j, t-1} / 2\right)
$$

\vspace{5mm}

\noindent where we can rewrite $\mathbf{W}_{j, t}$ as a discounted function of $\mathbf{C}_{j, t-1}$,   $\mathbf{W}_{j, t}=\mathbf{C}_{j, t-1}\left(1-\delta_{j}\right) / \delta_{j}$ for $0 < \delta_{j} \leq 1$ and the beta random variable $\eta_{j, t}$ is defined by the discount factor $0  < \kappa_{j} \leq 1$.  Discount methods are used to induce time-variations in the evolution of parameters and have been extensively used in many applications (\citealp{raftery2010online}, \citealp{dangl2012predictive}, \citealp{koop2013large}, \citealp{mcalinn2020multivariate} and others ) and well documented in \cite{prado2010time}. Note that lower values of $\delta$ and $\kappa$ induce higher degrees of variation in parameters and when discount factors are equal to one, both coefficients and volatilities will be constant.

Hence, the prior for time $t$ is given by 

\begin{equation}
\theta_{j t}, \sigma_{j t}^{-1} \mid \mathcal{D}_{t-1} \sim\mathcal{NG}\left(\boldsymbol{a_{j t}}, \boldsymbol{R_{j t}}, r_{j t}, r_{j t} s_{j, t-1}\right)
\end{equation}

\vspace{4mm}

\noindent where $r_{j t}=\kappa_{j} n_{j, t-1}$, $\boldsymbol{a_{j t}} = \boldsymbol{m_{j, t-1}}$ and  $\boldsymbol{R_{j t}} = \boldsymbol{C_{j, t-1}} / \delta_{j}$. 

\noindent \textbf{1-step ahead forecasts at time $t-1$}: The predictive distribution for t at time $t-1$ will be given by a student-t distribution with $r_{j t}$ degrees of freedom:

$$
y_{j t} \mid y_{<j t}, \mathcal{D}_{t-1} \sim \mathcal{T}_{r_{j t}}\left(f_{j t} ,   q_{j t}\right)
$$

\vspace{4mm}

\noindent with $f_{j t} = \mathbf{F}_{j t}^{\prime}\boldsymbol{a_{j t}}$ and $q_{j t} = s_{j, t-1}+\mathbf{F}_{j t}^{\prime} \mathbf{R}_{j t} \mathbf{F}_{j t} $. To make it explicit, we can define as the following manner

\vspace{4mm}

$$
\boldsymbol{a_{j t}} =\left(\begin{array}{c}
a_{j \beta t} \\
a_{j \gamma t}
\end{array}\right) \quad \text { and } \quad \mathbf{R}_{j t} =\left(\begin{array}{cc}
R_{j \beta t} & R_{j \beta \gamma t} \\
R_{j \beta \gamma t}^{\prime} & R_{j \gamma t}
\end{array}\right)
$$

we have

$$
\begin{array}{l}
f_{j t} = x_{j t-1}^{\prime} a_{j \beta t}+y_{< j t}^{\prime} a_{j \gamma t} \\
q_{j t} = s_{j, t-1}+y_{< j t}^{\prime} R_{j \gamma t} y_{< j t} + 2 y_{< j t}^{\prime} R_{j \beta \gamma t}^{\prime} x_{j t-1} + x_{j t-1}^{\prime} R_{j \beta t} x_{j t-1}
\end{array}
$$

\vspace{4mm}

\noindent \textbf{Updating at time t}: with the previous prior, the Normal-Gamma posterior is 

\begin{equation}
\theta_{j t}, \sigma_{j t}^{-1} \mid \mathcal{D}_{t} \sim\mathcal{NG}\left(\boldsymbol{m_{j t}}, \boldsymbol{C_{j t}}, n_{j t}, n_{j t} s_{j, t}\right)
\end{equation}

\vspace{4mm}

\noindent with parameters following standard updating equations:

\vspace{4mm}

Posterior mean vector: \quad \quad \quad \quad \quad \quad \quad \quad \quad $\boldsymbol{m_{j t}}=\boldsymbol{a_{j t}}+\mathbf{A}_{j t} e_{j t} $

Posterior covariance matrix factor: \quad  \quad \quad \quad $\boldsymbol{C_{j t}}=\left(\boldsymbol{R_{j t}}-\mathbf{A}_{j t} \mathbf{A}_{j t}^{\prime} q_{j t}\right) z_{j t} $

Posterior degrees of freedom: \quad \quad \quad \quad \quad \quad $n_{j t}=r_{j t}+1 $

Posterior residual variance estimate:  \quad \quad \quad $s_{j t}=s_{j, t-1} z_{j t} $

\vspace{4mm}
where 
\vspace{4mm}

1 - step ahead forecast error:  \quad  \quad \quad \quad \quad \quad$
e_{j t}=y_{j t}-\boldsymbol{F}_{j t}^{\prime} \boldsymbol{a_{j t}}
$

1-step ahead forecast variance factor: \quad \quad $q_{j t}=s_{j, t-1}+\boldsymbol{F_{j t}^{\prime}} \boldsymbol{R_{j t} F_{j t}}$ 

Adaptive coefficient vector: \quad \quad \quad \quad \quad \quad $\mathbf{A}_{j t}=\boldsymbol{R}_{j t} \boldsymbol{F}_{j t} / q_{j t}
$

Volatility update factor: \quad \quad \quad \quad \quad \quad \quad \quad $ z_{j t}=\left(r_{j t}+e_{j t}^{2} / q_{j t}\right) /\left(r_{j t}+1\right)
$

\clearpage

\section{Appendix B: Joint Predictive Moments}
\label{appendix B}

After computing the predictive density for each equation j, we are able to compute the joint predictive density for $\boldsymbol{y_{t}} $ conditional on the $\textit{parents}$:  ,

\begin{equation}
p\left(\mathbf{y}_{t} \mid \mathcal{D}_{t-1}\right)=\prod_{j=1: m} p\left(y_{j t} \mid \mathbf{y}_{< j t}, \mathcal{D}_{t-1}\right)
\end{equation}

\vspace{3mm}

\noindent being simply the product of the already computed m different univariate student-t distributions. Hence, after series being decoupled for sequential analysis, they are recoupled for multivariate forecasting. In our decision analysis at Section \ref{ref: Portfolio_section}, we are concerned with the mean and variance of this distribution for the portfolio allocation study:

\vspace{3mm}

\begin{equation}
\boldsymbol{f_{t}}=E\left(\mathbf{y_{t}} \mid \mathcal{D}_{t-1}\right), \quad \boldsymbol{Q_{t}}=V\left(\mathbf{y_{t}} \mid \mathcal{D}_{t-1}\right)
\end{equation}

\vspace{3mm}

The triangular form in Equation (2) allows for a recursive computation of moments according to the orderingdependence. Since the first dependent variable has a empty parental set, the forecast mean and variance for $\boldsymbol{j = 1}$ are given by 

\vspace{3mm}

$$
\begin{aligned}
f_{1t} &=x_{1 t-1}^{\prime} a_{1 \beta t} \\
q_{1t} &=\frac{r_{1 t}}{r_{1 t}-2}\left(x_{1, t-1}^{\prime} R_{1 \beta t} x_{1, t-1}+s_{1, t-1}\right)
\end{aligned}
$$

\vspace{3mm}

\noindent inserting $f_{1t}$ as the first element of $\boldsymbol{f_{t}}$ and $q_{1t} $ the $(1,1)$ element of $\boldsymbol{Q_{t}}$. For $\boldsymbol{j= 2, \ldots, m}$, we can find sequentially the subsequent predicted moments. Their conditional distributions also follow Student's t-distribution, with predictive moments given by

\vspace{3mm}

$$
\begin{aligned}
f_{jt} &=x_{j t-1}^{\prime} a_{j \beta t}+f_{< j t}^{\prime} a_{j \gamma t} \\
q_{jt} &=\frac{r_{j t}}{r_{j t}-2}\left(s_{j, t-1}+u_{j t}\right) +a_{j \gamma t}^{\prime} \boldsymbol{Q_{<j t}} a_{j y\gamma t}
\end{aligned}
$$

\vspace{3mm}

\noindent with $u_{j t}=f_{< j t}^{\prime} R_{j \gamma} f_{< j t}+\operatorname{tr}\left(R_{j \gamma t} \boldsymbol{Q_{<j t}}\right)+2 x_{j t-1}^{\prime} R_{j \beta \gamma} f_{< j t}+x_{j t-1}^{\prime} R_{j \beta t} x_{j t-1}$. Now, we just need to plug $f_{jt}$ as the j-th element of $\boldsymbol{f_{t}}$ and $q_{jt} $ the $(j,j)$ element of  $\boldsymbol{Q_{t}}$. Finally, the covariance vector among $y_{j t}$ and its $\textit{parents}$ $y_{< j t}$ is computed as $C\left(y_{j t}, y_{< j, t} \mid \mathcal{D}_{t-1}\right)=Q_{< j t} a_{j \gamma t} $. Hence, after reaching $j = m$, we have filled all elements of the $m$-vector $\boldsymbol{f_{t}}$ and the $m \times m$ covariance matrix $\boldsymbol{Q_{t}}$.


\section{Appendix C: Additional Results}
\label{appendix C}

In this Section we make several comparisons of the DOL approach in relation to different model settings. We show statistical and economic performances for models with only constant parameters (CP), time-varying parameters (TVP), constant volatilities (CV) and time-varying volatilties (SV).

\subsection{1. Portfolio Allocation}

In the Portfolio Allocation problem, we set $\delta = 0.995$ for TVP and $\kappa = 0.98$ for SV. The constant coefficient analogous are set equal to one. 

\vspace{6mm}

\subsubsection{1.1 Statistical Performance}

\vspace{6mm}

\begin{table}[!htbp] \centering 
  \caption{Statistical Performance Relative to W-RW-SV} 
  \label{} 
\begin{tabular}{@{\extracolsep{5pt}}l ccc} 
\\[-1.8ex]\hline 
\hline \\[-1.8ex] 
 & MSFE & LPDR \\ 
\hline \\[-1.8ex] 

DOA-CP-CV & $0.942$ & $-26.5$ \\ 
DOA-TVP-CV & $0.942$ & $-5.0$ \\ 
DOA-CP-SV & $0.938$ & $34.3$ \\ 
DOA-TVP-SV & $0.935$ & $52.2$ \\ 
&  &  \\ 
DOS-CP-CV & $0.951$ & $-38.6$ \\ 
DOS-TVP-CV & $0.944$ & $-9.7$ \\ 
DOS-CP-SV & $0.940$ & $23.4$ \\ 
DOS-TVP-SV & $0.945$ & $36.0$ \\ 
&  &  \\ 
W-RW-CV & $1.00$ & $-82.4$ \\ 
W-RW-SV & $1.00$ & $0$ \\ 
\hline \\[-1.8ex] 
\end{tabular} 
\begin{tablenotes}
      \small 
 \item       \textit{The table reports point (MSFE) and density (LPDR) out-of-sample forecasting metrics for different model settings compared to the Wishart Random-Walk Model with time-varying volatilities (W-RW-SV). CP, TVP, CV and SV are representing models with constant parameters, time-varying parameters, constant volatilities and time-varying volatilities, respectively. }
    \end{tablenotes}
\end{table}

\begin{figure}[!h]
\begin{center}
	\includegraphics[width=12cm]{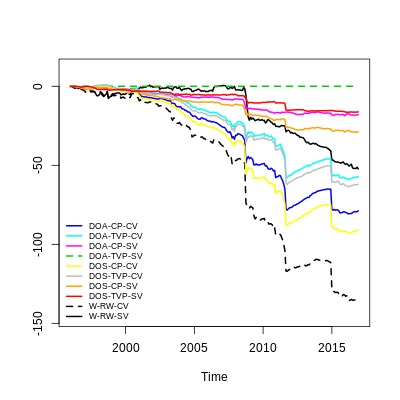} 
\caption{Accumulated Log Predictive Likelihood relative to the DOA-TVP-SV model.}
\label{fig: figura_appendix1}
\end{center}
\end{figure}


\clearpage

\subsubsection{1.2 Economic Performance}

\vspace{6mm}

\begin{table}[!htbp] \centering 
  \caption{Portfolio Performances} 
  \label{} 
\begin{tabular}{@{\extracolsep{5pt}}l ccccc} 
\\[-1.8ex]\hline 
\hline \\[-1.8ex] 
 & SR & $\Phi (\gamma=2)$ & $\Phi (\gamma=6)$ & $\Phi (\gamma=10)$ \\ 
\hline \\[-1.8ex] 

DOA-CP-CV & $1.32$ & $768$ & $764$ & $761$ \\ 
DOA-TVP-CV & $1.26$ & $758$ & $725$ & $691$ \\ 
DOA-CP-SV & $1.37$ & $718$ & $762$ & $810$ \\ 
DOA-TVP-SV & $1.38$ & $756$ & $788$ & $824$ \\ 
&  &  &  &  \\ 
DOS-CP-CV & $1.26$ & $659$ & $671$ & $684$ \\ 
DOS-TVP-CV & $1.24$ & $732$ & $696$ & $658$ \\ 
DOS-CP-SV & $1.34$ & $687$ & $728$ & $773$ \\ 
DOS-TVP-SV & $1.31$ & $650$ & $692$ & $739$ \\ 
&  &  &  &  \\ 
W-RW-CV & $0.56$ & $-139$ & $-155$ & $-171$ \\ 
W-RW-SV & $0.69$ & $0$ & $0$ & $0$ \\ 
\hline \\[-1.8ex] 
\end{tabular} 
\begin{tablenotes}
      \small 
 \item       \textit{The table reports out-of-sample portfolio performances in terms Sharpe Ratios (SR) and management fees ($\Phi$) for different model settings compared to the Wishart Random-Walk Model with time-varying volatilities (W-RW-SV). CP, TVP, CV and SV are representing models with constant parameters, time-varying parameters, constant volatilities and time-varying volatilities, respectively. We show management fees for different levels of relative risk aversion ($\gamma$).}
    \end{tablenotes}
\end{table}

\vspace{6mm}

\break

\subsection{2. Macroeconomic Forecasting}

In the Macroeconomic Forecasting problem, we use $\delta = 0.99$ for TVP and $\kappa = 0.96$ for SV. The constant coefficient analogous are set equal to one.

\vspace{6mm}

\begin{table}[!htbp] \centering 
  \caption{Statistical Performance Relative to DOA-TVP-SV} 
  \label{} 
\begin{tabular}{@{\extracolsep{5pt}}l ccc} 
\\[-1.8ex]\hline 
\hline \\[-1.8ex] 
 & MSFE & LPDR \\ 
\hline \\[-1.8ex] 
DOA-CP-CV & $1.12$ & $$-$39.8$ \\ 
DOA-TVP-CV & $1.07$ & $$-$35.2$ \\ 
DOA-CP-SV & $1.05$ & $$-$7.8$ \\ 
DOA-TVP-SV & $1.00$ & $0.0$ \\ 
DOS-CP-CV & $1.09$ & $$-$38.160$ \\ 
DOS-TVP-CV & $1.10$ & $$-$34.3$ \\ 
DOS-CP-SV & $1.05$ & $$-$11.4$ \\ 
DOS-TVP-SV & $0.98$ & $$-$3.5$ \\ 
\hline \\[-1.8ex] 
\end{tabular} 
\begin{tablenotes}
      \small 
 \item       \textit{The table reports point (MSFE) and density (LPDR) out-of-sample forecasting metrics for different model settings compared to the DOA-TVP-SV, where CP, TVP, CV and SV are representing models with constant parameters, time-varying parameters, constant volatilities and time-varying volatilities, respectively. }
    \end{tablenotes}
\end{table}

\begin{figure}[!h]
\begin{center}
	\includegraphics[width=11cm]{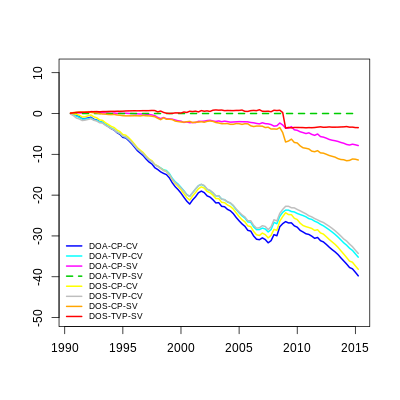} 
\caption{Accumulated Log Predictive Likelihood relative to the DOA-TVP-SV}
\label{fig: figura_appendix2}
\end{center}
\end{figure}

\vspace{4mm}

\clearpage

\bibliographystyle{ecta}
\bibliography{referencias}

\begin{thebibliography}{46}
\newcommand{\enquote}[1]{``#1''}
\expandafter\ifx\csname natexlab\endcsname\relax\def\natexlab#1{#1}\fi

\bibitem[\protect\citeauthoryear{Beckmann, Koop, Korobilis, and
  Sch{\"u}ssler}{Beckmann et~al.}{2020}]{beckmann2020exchange}
\textsc{Beckmann, J., G.~Koop, D.~Korobilis, and R.~A. Sch{\"u}ssler} (2020):
  \enquote{Exchange rate predictability and dynamic Bayesian learning,}
  \emph{Journal of Applied Econometrics}, 35, 410--421.

\bibitem[\protect\citeauthoryear{Byrne, Korobilis, and Ribeiro}{Byrne
  et~al.}{2016}]{byrne2016exchange}
\textsc{Byrne, J.~P., D.~Korobilis, and P.~J. Ribeiro} (2016):
  \enquote{Exchange rate predictability in a changing world,} \emph{Journal of
  International Money and Finance}, 62, 1--24.

\bibitem[\protect\citeauthoryear{Byrne, Korobilis, and Ribeiro}{Byrne
  et~al.}{2018}]{byrne2018sources}
---\hspace{-.1pt}---\hspace{-.1pt}--- (2018): \enquote{On the sources of
  uncertainty in exchange rate predictability,} \emph{International Economic
  Review}, 59, 329--357.

\bibitem[\protect\citeauthoryear{Catania, Grassi, and Ravazzolo}{Catania
  et~al.}{2019}]{catania2019forecasting}
\textsc{Catania, L., S.~Grassi, and F.~Ravazzolo} (2019): \enquote{Forecasting
  cryptocurrencies under model and parameter instability,} \emph{International
  Journal of Forecasting}, 35, 485--501.

\bibitem[\protect\citeauthoryear{Cenesizoglu and Timmermann}{Cenesizoglu and
  Timmermann}{2012}]{cenesizoglu2012return}
\textsc{Cenesizoglu, T. and A.~Timmermann} (2012): \enquote{Do return
  prediction models add economic value?} \emph{Journal of Banking \& Finance},
  36, 2974--2987.

\bibitem[\protect\citeauthoryear{Clark and McCracken}{Clark and
  McCracken}{2010}]{clark2010averaging}
\textsc{Clark, T.~E. and M.~W. McCracken} (2010): \enquote{Averaging forecasts
  from VARs with uncertain instabilities,} \emph{Journal of Applied
  Econometrics}, 25, 5--29.

\bibitem[\protect\citeauthoryear{Cogley and Sargent}{Cogley and
  Sargent}{2005}]{cogley2005drifts}
\textsc{Cogley, T. and T.~J. Sargent} (2005): \enquote{Drifts and volatilities:
  monetary policies and outcomes in the post WWII US,} \emph{Review of Economic
  dynamics}, 8, 262--302.

\bibitem[\protect\citeauthoryear{Costa, Smith, Nichols, Cussens, Duff, Makin
  et~al.}{Costa et~al.}{2015}]{costa2015searching}
\textsc{Costa, L., J.~Smith, T.~Nichols, J.~Cussens, E.~P. Duff, T.~R. Makin,
  et~al.} (2015): \enquote{Searching multiregression dynamic models of
  resting-state fMRI networks using integer programming,} \emph{Bayesian
  Analysis}, 10, 441--478.

\bibitem[\protect\citeauthoryear{Dangl and Halling}{Dangl and
  Halling}{2012}]{dangl2012predictive}
\textsc{Dangl, T. and M.~Halling} (2012): \enquote{Predictive regressions with
  time-varying coefficients,} \emph{Journal of Financial Economics}, 106,
  157--181.

\bibitem[\protect\citeauthoryear{Del~Negro and Primiceri}{Del~Negro and
  Primiceri}{2015}]{del2015time}
\textsc{Del~Negro, M. and G.~E. Primiceri} (2015): \enquote{Time varying
  structural vector autoregressions and monetary policy: a corrigendum,}
  \emph{The review of economic studies}, 82, 1342--1345.

\bibitem[\protect\citeauthoryear{Della~Corte, Sarno, and Tsiakas}{Della~Corte
  et~al.}{2009}]{della2009economic}
\textsc{Della~Corte, P., L.~Sarno, and I.~Tsiakas} (2009): \enquote{An economic
  evaluation of empirical exchange rate models,} \emph{The review of financial
  studies}, 22, 3491--3530.

\bibitem[\protect\citeauthoryear{Della~Corte and Tsiakas}{Della~Corte and
  Tsiakas}{2012}]{della2012statistical}
\textsc{Della~Corte, P. and I.~Tsiakas} (2012): \enquote{Statistical and
  economic methods for evaluating exchange rate predictability,} \emph{Handbook
  of exchange rates}, 221--263.

\bibitem[\protect\citeauthoryear{Fisher, Pettenuzzo, Carvalho et~al.}{Fisher
  et~al.}{2020}]{fisher2020optimal}
\textsc{Fisher, J.~D., D.~Pettenuzzo, C.~M. Carvalho, et~al.} (2020):
  \enquote{Optimal asset allocation with multivariate Bayesian dynamic linear
  models,} \emph{Annals of Applied Statistics}, 14, 299--338.

\bibitem[\protect\citeauthoryear{Fleming, Kirby, and Ostdiek}{Fleming
  et~al.}{2001}]{fleming2001economic}
\textsc{Fleming, J., C.~Kirby, and B.~Ostdiek} (2001): \enquote{The economic
  value of volatility timing,} \emph{The Journal of Finance}, 56, 329--352.

\bibitem[\protect\citeauthoryear{Gamerman and Lopes}{Gamerman and
  Lopes}{2006}]{lopes2006mcmc}
\textsc{Gamerman, D. and H.~F. Lopes} (2006): \enquote{MCMC-Stochastic
  Simulation for Bayesian Inference,} \emph{Chapman Hill}.

\bibitem[\protect\citeauthoryear{Gruber and West}{Gruber and
  West}{2016}]{gruber2016gpu}
\textsc{Gruber, L. and M.~West} (2016): \enquote{GPU-accelerated Bayesian
  learning and forecasting in simultaneous graphical dynamic linear models,}
  \emph{Bayesian Analysis}, 11, 125--149.

\bibitem[\protect\citeauthoryear{Hoeting, Madigan, Raftery, and
  Volinsky}{Hoeting et~al.}{1999}]{hoeting1999bayesian}
\textsc{Hoeting, J.~A., D.~Madigan, A.~E. Raftery, and C.~T. Volinsky} (1999):
  \enquote{Bayesian model averaging: a tutorial,} \emph{Statistical science},
  382--401.

\bibitem[\protect\citeauthoryear{Jegadeesh and Titman}{Jegadeesh and
  Titman}{1993}]{jegadeesh1993returns}
\textsc{Jegadeesh, N. and S.~Titman} (1993): \enquote{Returns to buying winners
  and selling losers: Implications for stock market efficiency,} \emph{The
  Journal of finance}, 48, 65--91.

\bibitem[\protect\citeauthoryear{Kang, Xie, and Wang}{Kang
  et~al.}{2020}]{kang2020cholesky}
\textsc{Kang, X., C.~Xie, and M.~Wang} (2020): \enquote{A Cholesky-based
  estimation for large-dimensional covariance matrices,} \emph{Journal of
  Applied Statistics}, 47, 1017--1030.

\bibitem[\protect\citeauthoryear{Kastner and Huber}{Kastner and
  Huber}{2020}]{kastner2020sparse}
\textsc{Kastner, G. and F.~Huber} (2020): \enquote{Sparse Bayesian vector
  autoregressions in huge dimensions,} \emph{Journal of Forecasting}.

\bibitem[\protect\citeauthoryear{Koop and Korobilis}{Koop and
  Korobilis}{2012}]{koop2012forecasting}
\textsc{Koop, G. and D.~Korobilis} (2012): \enquote{Forecasting inflation using
  dynamic model averaging,} \emph{International Economic Review}, 53, 867--886.

\bibitem[\protect\citeauthoryear{Koop and Korobilis}{Koop and
  Korobilis}{2013}]{koop2013large}
---\hspace{-.1pt}---\hspace{-.1pt}--- (2013): \enquote{Large time-varying
  parameter VARs,} \emph{Journal of Econometrics}, 177, 185--198.

\bibitem[\protect\citeauthoryear{Koop and Korobilis}{Koop and
  Korobilis}{2014}]{koop2014new}
---\hspace{-.1pt}---\hspace{-.1pt}--- (2014): \enquote{A new index of financial
  conditions,} \emph{European Economic Review}, 71, 101--116.

\bibitem[\protect\citeauthoryear{Krueger}{Krueger}{2015}]{krueger2015bvarsv}
\textsc{Krueger, F.} (2015): \enquote{bvarsv: Bayesian Analysis of a Vector
  Autoregressive Model with Stochastic Volatility and Time-Varying Parameters,}
  \emph{R package: cran. r-project. org/package= bvarsv}.

\bibitem[\protect\citeauthoryear{Lavine, Lindon, West et~al.}{Lavine
  et~al.}{2020}]{lavine2020adaptive}
\textsc{Lavine, I., M.~Lindon, M.~West, et~al.} (2020): \enquote{Adaptive
  variable selection for sequential prediction in multivariate dynamic models,}
  \emph{Bayesian Analysis}.

\bibitem[\protect\citeauthoryear{Levy and Lopes}{Levy and
  Lopes}{2021}]{levy2021time}
\textsc{Levy, B.~P. and H.~F. Lopes} (2021): \enquote{Trend-Following
  Strategies via Dynamic Momentum Learning,} \emph{arXiv preprint
  arXiv:2106.08420}.

\bibitem[\protect\citeauthoryear{Litterman}{Litterman}{1986}]{litterman1986forecasting}
\textsc{Litterman, R.~B.} (1986): \enquote{Forecasting with Bayesian vector
  autoregressions—five years of experience,} \emph{Journal of Business \&
  Economic Statistics}, 4, 25--38.

\bibitem[\protect\citeauthoryear{Lopes, McCulloch, and Tsay}{Lopes
  et~al.}{2018}]{lopes2016parsimony}
\textsc{Lopes, H.~F., R.~E. McCulloch, and R.~S. Tsay} (2018):
  \enquote{Parsimony inducing priors for large scale state-space models,}
  \emph{Technical Report 2018-08}.

\bibitem[\protect\citeauthoryear{Madigan and Raftery}{Madigan and
  Raftery}{1994}]{madigan1994model}
\textsc{Madigan, D. and A.~E. Raftery} (1994): \enquote{Model selection and
  accounting for model uncertainty in graphical models using Occam's window,}
  \emph{Journal of the American Statistical Association}, 89, 1535--1546.

\bibitem[\protect\citeauthoryear{Marquering and Verbeek}{Marquering and
  Verbeek}{2004}]{marquering2004economic}
\textsc{Marquering, W. and M.~Verbeek} (2004): \enquote{The economic value of
  predicting stock index returns and volatility,} \emph{Journal of Financial
  and Quantitative Analysis}, 39, 407--429.

\bibitem[\protect\citeauthoryear{McAlinn, Aastveit, Nakajima, and West}{McAlinn
  et~al.}{2020}]{mcalinn2020multivariate}
\textsc{McAlinn, K., K.~A. Aastveit, J.~Nakajima, and M.~West} (2020):
  \enquote{Multivariate Bayesian predictive synthesis in macroeconomic
  forecasting,} \emph{Journal of the American Statistical Association}, 115,
  1092--1110.

\bibitem[\protect\citeauthoryear{McAlinn and West}{McAlinn and
  West}{2019}]{mcalinn2019dynamic}
\textsc{McAlinn, K. and M.~West} (2019): \enquote{Dynamic Bayesian predictive
  synthesis in time series forecasting,} \emph{Journal of econometrics}, 210,
  155--169.

\bibitem[\protect\citeauthoryear{Meese and Rogoff}{Meese and
  Rogoff}{1983}]{meese1983empirical}
\textsc{Meese, R.~A. and K.~Rogoff} (1983): \enquote{Empirical exchange rate
  models of the seventies: Do they fit out of sample?} \emph{Journal of
  international economics}, 14, 3--24.

\bibitem[\protect\citeauthoryear{Moskowitz, Ooi, and Pedersen}{Moskowitz
  et~al.}{2012}]{moskowitz2012time}
\textsc{Moskowitz, T.~J., Y.~H. Ooi, and L.~H. Pedersen} (2012): \enquote{Time
  series momentum,} \emph{Journal of financial economics}, 104, 228--250.

\bibitem[\protect\citeauthoryear{Nakajima and West}{Nakajima and
  West}{2013}]{nakajima2013bayesian}
\textsc{Nakajima, J. and M.~West} (2013): \enquote{Bayesian analysis of latent
  threshold dynamic models,} \emph{Journal of Business \& Economic Statistics},
  31, 151--164.

\bibitem[\protect\citeauthoryear{Prado and West}{Prado and
  West}{2010}]{prado2010time}
\textsc{Prado, R. and M.~West} (2010): \emph{Time series: modeling,
  computation, and inference}, CRC Press.

\bibitem[\protect\citeauthoryear{Primiceri}{Primiceri}{2005}]{primiceri2005time}
\textsc{Primiceri, G.~E.} (2005): \enquote{Time varying structural vector
  autoregressions and monetary policy,} \emph{The Review of Economic Studies},
  72, 821--852.

\bibitem[\protect\citeauthoryear{Queen, Wright, and Albers}{Queen
  et~al.}{2008}]{queen2008forecast}
\textsc{Queen, C.~M., B.~J. Wright, and C.~J. Albers} (2008): \enquote{Forecast
  covariances in the linear multiregression dynamic model,} \emph{Journal of
  Forecasting}, 27, 175--191.

\bibitem[\protect\citeauthoryear{Raftery, K{\'a}rn{\`y}, and Ettler}{Raftery
  et~al.}{2010}]{raftery2010online}
\textsc{Raftery, A.~E., M.~K{\'a}rn{\`y}, and P.~Ettler} (2010):
  \enquote{Online prediction under model uncertainty via dynamic model
  averaging: Application to a cold rolling mill,} \emph{Technometrics}, 52,
  52--66.

\bibitem[\protect\citeauthoryear{Rossi}{Rossi}{2013}]{rossi2013exchange}
\textsc{Rossi, B.} (2013): \enquote{Exchange rate predictability,}
  \emph{Journal of economic literature}, 51, 1063--1119.

\bibitem[\protect\citeauthoryear{Shirota, Omori, Lopes, and Piao}{Shirota
  et~al.}{2017}]{shirota2017cholesky}
\textsc{Shirota, S., Y.~Omori, H.~F. Lopes, and H.~Piao} (2017):
  \enquote{Cholesky realized stochastic volatility model,} \emph{Econometrics
  and Statistics}, 3, 34--59.

\bibitem[\protect\citeauthoryear{Sims}{Sims}{1980}]{sims1980macroeconomics}
\textsc{Sims, C.~A.} (1980): \enquote{Macroeconomics and reality,}
  \emph{Econometrica: journal of the Econometric Society}, 1--48.

\bibitem[\protect\citeauthoryear{West}{West}{2020}]{west2020bayesian}
\textsc{West, M.} (2020): \enquote{Bayesian forecasting of multivariate time
  series: scalability, structure uncertainty and decisions,} \emph{Annals of
  the Institute of Statistical Mathematics}, 72, 1--31.

\bibitem[\protect\citeauthoryear{West and Harrison}{West and
  Harrison}{1997}]{west2006bayesian}
\textsc{West, M. and J.~Harrison} (1997): \emph{Bayesian forecasting and
  dynamic models}, Springer Science \& Business Media.

\bibitem[\protect\citeauthoryear{Zhao, Xie, and West}{Zhao
  et~al.}{2016}]{zhao2016dynamic}
\textsc{Zhao, Z.~Y., M.~Xie, and M.~West} (2016): \enquote{Dynamic dependence
  networks: Financial time series forecasting and portfolio decisions,}
  \emph{Applied Stochastic Models in Business and Industry}, 32, 311--332.

\bibitem[\protect\citeauthoryear{Zheng, Tsui, Kang, and Deng}{Zheng
  et~al.}{2017}]{zheng2017cholesky}
\textsc{Zheng, H., K.-W. Tsui, X.~Kang, and X.~Deng} (2017):
  \enquote{Cholesky-based model averaging for covariance matrix estimation,}
  \emph{Statistical Theory and Related Fields}, 1, 48--58.

\end{thebibliography}
\clearpage




\end{document}